\documentclass[pre,graphicx,preprint,epsf,epsfig,amsmath,letterpaper]{revtex4}
\usepackage{graphicx,pdflscape}% Include figure files
\usepackage{alltt,dsfont}
\newcommand{\figurewidth}{9cm}

\newcommand{\llabel}[1]{\label{#1} }
\newcommand{\ra}{\rightarrow}

\newcommand{\beq}{\begin{equation}}
\newcommand{\eeq}{\end{equation}}
\newcommand{\barray}{\begin{eqnarray}}
\newcommand{\earray}{\end{eqnarray}}

\newcommand{\nn}{\nonumber}

\begin{document}
\title{Dynamics of Energy Transport in a Toda Ring}
\author{B.~Sriram Shastry and A.~P.~Young  }
\affiliation{Physics Department, University of California, Santa Cruz, CA 95064 }
\date{\today}
\begin{abstract}
We present results on the relationships between persistent currents and the
known conservation laws in the classical Toda ring. We also show that
perturbing  the integrability leads to a decay of the  currents at long times,
with a time scale that is determined by the perturbing parameter. We summarize
several known results concerning the Toda ring in 1-dimension, and present
new results relating to the frequency, average kinetic and potential energy, and mean square
displacement in the cnoidal waves, as functions of the wave vector and a
parameter that determines the non linearity.  
\end{abstract}
\maketitle

\section{Introduction}

\llabel{intro}

Toda's non linear lattice~\cite{toda} is one of the very few examples of non
linear lattices in condensed matter physics, where explicit analytical
solutions are available for the dynamics. 
There are several aspects of condensed matter physics where the Toda lattice
is a useful model. Toda himself applied his non linear lattice to understand
heat propagation~\cite{toda_2}, and further studies with added impurities throw
interesting light on this phenomenon~\cite{sataric}.  Interestingly, the
lattice has also found recent applications in the  context of the dynamics of
DNA~\cite{muto}, where the Toda interaction is a reasonable representation of
the known non-linear couplings between base pairs. It has also been used
to represent the potential of hydrogen and peptide bonds in the
$\alpha$-helix~\cite{ovidio}.

In this paper we study in detail the Toda lattice with periodic boundary
conditions (the Toda ring).
Our aim is two-fold. Firstly we derive several \textit{new}
results.
Secondly, since the Toda lattice and its properties
are less well known to students of condensed matter than they deserve to be,
we collect
together some of the basics of the model and its solution in a
form and notation that is standard in condensed matter physics. 

The excitations of this lattice are
not phonons as in a harmonic lattice, but can be expressed in terms of non
linear excitations that are termed solitonic. For periodic boundary
conditions the excitations are more properly the {\em cnoidal
waves} corresponding to a family of waves characterized by a wave vector and
another parameter, related to the non linearity of the excitations,
namely the elliptic parameter $m$ discussed below in
Sec.~\ref{harm_anharm}.
The
lattice is very simple to describe and the solution is both beautiful and
instructive. Surprisingly, we find that the dispersion relation of the
excitations
given by Toda is only correct in the limit of weak anharmonicity or long
wavelength. Here, we give a complete expression, which does not seem to have been
calculated before.

Soon after Toda found the exact solution, his model
was found to be exactly
integrable~\cite{henon,flaschka}, i.e.~it has an infinite set of ``generalized
conservation laws''. These conservation laws are expressed
through conserved currents that Poisson commute with the Hamiltonian as well as each other, and the
stability of the solitons is understood to arise from the existence of these
currents.

There is considerable interest in the role of the conservation laws
in the transport of heat or energy, and the Toda lattice provides an excellent
model to test some ideas about their role in transport, as detailed below.
Since the model is classical we can study reasonably large systems (up to 64
atoms in our largest studies), unlike in quantum systems where the Hilbert
space grows exponentially with the number of sites.

We address two specific
issues in transport theory in this paper.
The first
is the  role of  integrability of the model in determining the exact value of
the asymptotic correlation function of the energy current---this value provides
us with the coefficient of the delta function in the thermal conductivity at
zero frequency. It is known as the Drude term in the Kubo conductivity, and
is widely discussed
in current literature.  The second issue concerns the role
of perturbations of integrable models, whereby conservation laws are
destroyed.  We  present results on the decay of the energy current in a
slightly perturbed Toda lattice for various values of the parameter that
destroys integrability, and show that there is an underlying scaling picture
which provides a general understanding of this phenomenon.

The plan of the paper is as follows. In Sec.~\ref{model} A we
define the Toda ring, and in Sec.~\ref{model} B we discuss the extreme
limits of harmonic and anharmonic interactions. The frequency of periodic
solutions (cnoidal waves) of the Toda ring are derived in Sec.~\ref{model} C.
In Sec.~\ref{model} D we highlight the differences between the results for the
frequency spectrum derived in Sec.~\ref{model} C and the spectrum determined
by Toda. We also show how, in the extreme anharmonic limit and at long
wavelength, the cnoidal waves
can be viewed as a train of isolated solitons. 
Next, in 
Sec.~\ref{energy}, we calculate the
kinetic energy in the modes as a function of wave vector and 
anharmonicity parameter, and also the ratio of the kinetic energy and the
average displacement as functions of wave vector and anharmonicity.
In Sec.~\ref{conserved}
we list the conserved currents of the Toda ring obtained from the Lax matrix.
Next,
in Sec.~\ref{persistence}, we analyze the
persistence of energy currents by expanding the energy current in terms of
these conserved currents. Comparing with numerical results we show that the
persistent part of the conserved energy cannot be expressed in terms of the
Lax currents alone, but quadratic combinations of Lax currents are also needed
to get an accurate description. In Sec.~\ref{perturb}, we
consider a Toda ring in which a small interaction is added which breaks
integrability. We
calculate numerically the decay of the persistent currents as functions of time for
different values of the
perturbing parameter. We show that the results fit a scaling property that
is expected, but with substantial corrections to scaling
so very long runs are needed
to reach the asymptotic scaling regime. Finally, in Sec.~\ref{conclusions},
we summarize the main results of the paper and comment on them.
Appendix \ref{app-elliptic} contains some needed results on elliptic functions.
In Appendix \ref{app-toda} we discuss Toda's result for
the frequency spectrum $\omega^{T}_{k}$ of periodic waves, and explain why
this result is only correct in some limiting cases. Appendix \ref{alt-deriv}
gives an
alternative derivation of the dispersion relation of the cnoidal waves, while
Appendix \ref{potential-energy} computes the potential energy of the Toda ring.

As discussed above, one of the objectives of this paper is to collect results
on the Toda lattice in a form accessible to condensed matter physicists. In
addition, our
result for the dispersion of cnoidal waves in Eq.~(\ref{disp-ring}) has not been
published before, and the fact that Eq.~(\ref{disp-ring}) differs from
the result given by Toda, is new.
Other new results are the computation of the average potential and kinetic energies
in the cnoidal waves in Section \ref{energy}, and the expansion in Eq.~(\ref{toda-solution}).
Finally, our results for the connection between
persistent energy currents and the Lax conserved currents in
Sec.~\ref{persistence}, and
for the decay of the perturbed Toda ring in Sec.~\ref{perturb},
are new.

\section{Toda Ring, Cnoidal waves and their spectrum}

\label{model}
We summarize some interesting facts about the Toda lattice in this
section. By imposing periodic boundary conditions, we deal with a ring
of finite extent. The Hamiltonian of the Toda lattice is
\begin{equation}
H=\sum_{n=1}^{N} \frac{p_n^2}{2 M} + \frac{a}{b}
\sum_{n=1}^{N} \{ e^{-b(u_{n+1}-u_{n})} -1+ b (u_{n+1}-u_{n} ) \}.
\llabel{toda-H} 
\end{equation}
The displacement variable $u_n$ is
defined through  $R_n=R^0_n+ u_n$,  where $R^0_n$ is the equilibrium
position of the $n^{th}$ atom.   In principle the displacement $u_n$
ranges between $\pm \infty$, so for consistency, we must either imagine that
the  lattice constant $R^0_{n+1}-R^0_{n}$  is infinite as well,
or that the displacements
are transverse to the ring. Further
we assume   $u_n=u_{n+N}$ as appropriate for a ring geometry.  The two
non-exponential terms in the interaction potential are
irrelevant, since they add but a constant to the energy, but it is convenient
to include them 
since the ``two body potential'' then
explicitly displays a minimum at zero relative displacement.
The variable $p_n$ is conjugate to $u_n$ satisfying the standard Poisson
Bracket (PB) relation $\{ u_n,p_m\}= \delta_{n,m}$, and $a, b, M$ are
parameters. 

The equations of motion follow from Hamilton's equations,
\begin{align}
\dot{u}_n&= \frac{p_n}{M}, & \dot{p}_n& = - a
(e^{-b(u_{n+1}-u_{n})}-e^{-b(u_{n}-u_{n-1})}).\llabel{eom-1}
\end{align}
We note that the total momentum $p_\text{total}=\sum_n p_n$ is a constant of
motion.   The dynamics has in fact many more conservation laws,
a consequence of the property of integrability which was proved by
Henon~\cite{henon} and Flaschka~\cite{flaschka} for this system. The
explicit form of the conservation laws are given later in the paper, and
follow from the Lax structure that underlies the dynamics.  The
parameters $a, b,$ and $M$ give us explicit freedom to interpolate
between the harmonic and extremely anharmonic limits. 

\subsection{Harmonic and Anharmonic Limits}

\label{harm_anharm}
A formal Taylor expansion of  the  exponential interaction gives us
\begin{equation}
  H= \sum_{n=1}^{N} \frac{p_n^2}{2 M} + \frac{\kappa}{2}
  \sum_{n=1}^{N} \{ (u_{n+1}-u_{n})^2 - \frac{b}{3}(u_{n+1}-u_{n})^3+ 
  \frac{b^2}{12}(u_{n+1}-u_{n})^4 + \ldots \}, \llabel{toda-H-harmonic}
\end{equation} 
where $\kappa = a b$. As long as the displacements satisfy
$|u_j| b \ll1$, the
anharmonic terms do not become important, so we expect a harmonic
response. However for $|u_j| b \gg1 $, the anharmonic terms will dominate.
To recover the harmonic lattice,
one could formally take a limit  $b\rightarrow 0$ and simultaneously let
$a\rightarrow \kappa /b$ with $\kappa$ remaining finite,
so that the anharmonic terms are explicitly killed,

We will see below that the cnoidal waves of Toda, contain  a parameter, the
elliptic ``m'' parameter, which controls the amplitude of the waves, and
varying this parameter gives harmonic as well as strongly anharmonic response.
Physically, the elliptic parameter ``m'' may be  viewed as tuning the
anharmonicity, with the harmonic limit being $m \to 0$ and the extreme
anharmonic limit being $m \to 1$.
Mathematically, the parameter $m$ plays a fundamental role in
the theory of Jacobian elliptic functions\cite{ww}. 

For later use, we note that the harmonic limit has  a dispersion $\omega_k$
and sound velocity $c_0$ given by
\barray
\omega_k & =& c_0 \  2 |\sin{\frac{k}{2}}|, \nn \\
c_0&= & \sqrt{\frac{ a b}{M}}. \llabel{harmonic-1}
\earray

\subsection{Single parameter formulation} 

The Toda lattice is also integrable in quantum theory. 
Quantum integrability was established by Sutherland~\cite{bill},
Gutzwiller~\cite{gutz}, Sklyanin~\cite{sklyanin}, and Pasquier and
Gaudin~\cite{pasquier} using different formulations which are
summarized in the work of Siddharthan and Shastry~\cite{rsidd}, who also
establish their
equivalence.  In the viewpoint of ~\cite{bill},
developed by~\cite{rsidd}, the Toda lattice emerges from a
crystallization of a gas of impenetrable particles with an interaction
$\propto \{ \sinh(R_n-R_m)\}^{-2}$.

Quantum mechanically, it is possible to reduce the Toda lattice to a single parameter
problem~\cite{rsidd}. We scale $u_n \ra b \ u_n, \ p_n \ra \frac{1}{b} \
p_n, \ H\ra \frac{1}{ M b}  H$ and set $\eta=2 M a$, so that
\begin{equation}
H=\sum_{n=1}^{N} \frac{p_n^2}{2 } + \frac{\eta}{2} \sum_{n=1}^{N} \{
e^{-(u_{n+1}-u_{n})} -1+  (u_{n+1}-u_{n} ) \}. \llabel{toda-H-2}
\end{equation}
In this representation $\eta \ra \infty$ gives the harmonic
limit, since the displacements become very small so the potential energy
remains small. On the other hand, $\eta \ra 0$ corresponds to the extreme
anharmonic limit,  since now  the displacements are large, and hence 
high order terms in the expansion of the exponential matter, and ultimately
dominate. We need to keep in mind that in this
extreme non-linear limit of the model,
the particles are not allowed to cross so
they act as impenetrable billiard balls\cite{rsidd} with free propagation between
successive collisions. 
Classically, even this one parameter
$\eta$ can be removed by a rescaling of the displacements, so large
anharmonicity corresponds to large displacements and vice versa.

\subsection{Cnoidal wave solutions}
\label{cnoidal}
We derive here the formulas for the excitation spectrum of the Toda ring.
The relation with Toda's work is discussed in the Appendix \ref{app-toda}.
His  papers\cite{toda,toda_2} and book\cite{toda-book} focus on a set of dual
variables and give a solution for the displacement,
but we point out that his
dispersion relation is not appropriate for the periodic boundary conditions
(i.e.~a ring)
which is the focus of this paper.
Here we present an explicit solution for this case, which
is not available in
literature as far as we can tell\cite{bill-unpub}. 
Our Eq.~(\ref{ratio}) below, concerning the total energy to mean square
amplitude ratio, and related results, also seem to be new.  We feel that they,
too, are helpful in appreciating the Toda system from a condensed matter point
of view. 

The Toda Hamiltonian in Eq.~(\ref{toda-H}) leads to
the following equation of motion for the displacement\footnote{The mass is denoted by M rather than m, to avoid confusion with the Elliptic function parameter.}
:
\begin{equation}
b\ \ddot{u}_n= \frac{a b}{M} \ \{e^{-b(u_{n}-u_{n-1})}
- e^{-b(u_{n+1}-u_{n})} \}. \llabel{eom-2}
\end{equation}
We seek a special kind of solution namely a {\em constant profile solution}
\begin{align}
b \ u_j(t)& =  d_k( \phi_j(t)  ),  && \nn \\
\phi_j(t)&=  k \ j - \omega_k t =  k (j -  c_k t ) =
2 \pi \left( \frac{j}{\lambda}- \nu t \right) \llabel{const-profile-1},
\end{align}
where $\phi_j(t)$ is the usual phase factor depending linearly on space
and time  with wave vector $k$ (or equivalently
wave length $\lambda$), and angular
frequency $\omega_k$ (or equivalently frequency $\nu$), and we have defined the 
velocity by $c_k=\omega_k/|k|$. We will often omit
the argument of the phase for brevity. We set  
\begin{equation}
k= \frac{2 \pi}{N} \nu,
\label{kvalues}
\end{equation}
where the  $N$  integers $\nu$ obey  $1 \leq \nu \leq N$. Consequently,
the Brillouin Zone is the range $0 \leq k \leq 2 \pi$.
Periodic boundary
conditions, $u_n=u_{n+N},$ are satisfied if the function $d_k$ is periodic,
i.e.~ $d_k(x)=d_k(x+ 2 \pi)$. Here $\omega_k$ is yet to be determined, along
with the form of $d_k$. We now define a scaled
frequency $\bar{\omega}_k$ by $\omega_k = \sqrt{\frac{a b}{M}} \
\bar{\omega}_k$ so the equation of motion, Eq.~(\ref{eom-2}),
reduces to a non-linear,
differential, difference equation;
\begin{equation}
\bar{\omega}_k^2 \  d_k''(\phi)=
e^{ d_k(\phi-k)-d_k(\phi)}- e^{ d_k(\phi)-d_k(\phi+k)}. \llabel{eom-3}
\end{equation}
This equation is satisfied by the choice
\begin{align}
d_k(\phi)&=
\log\left\{\frac{\theta_4(\frac{\phi- k}{2})}{\theta_4(\frac{\phi}{2})}
\right\}, \label{sol-periodic}
\end{align}
where we summarize,
in Appendix \ref{app-elliptic}, the necessary definitions of the elliptic theta
functions as needed for this work~\footnote{We will always denote the elliptic
functions by the parameters $m = k^2$ and $m'=1-m=1-k^2=k'^2$ rather than by
their modulus $k,k'$ (in order to avoid confusion with the symbol for the wave
vector). \llabel{footnote-k-m}}. 
To see that Eq.~(\ref{sol-periodic}) solves Eq.~(\ref{eom-3})
we use the Jacobi addition formula for the $\theta$ functions~\cite{ww-1}, 
and write the first term in the RHS of Eq.~(\ref{eom-3}) as
\begin{eqnarray}
e^{ d_k(\phi-k)-d_k(\phi)}&=& \frac{\theta_4(\frac{\phi- 2 k}{2})
\theta_4(\frac{\phi}{2}) }{\theta^2_4(\frac{\phi-  k}{2})} \nn \\
&=& \frac{1}{\theta^2_4(0)} \left\{ \theta^2_4(\frac{k}{2}) -
\theta^2_1(\frac{k}{2}) \ \frac{\theta^2_1(\frac{\phi-k}{2})}
{\theta^2_4(\frac{\phi-k}{2})} \right\} .
\llabel{pot-1}
\end{eqnarray}
The second term in the RHS of Eq.~(\ref{eom-3}) is obtained by replacing
$\phi$ by $\phi + k$.
Upon using 
 the relationship between the theta functions and the Jacobian elliptic functions~\cite{gr-1} 
\begin{equation}
\frac{\theta_1(x)}{\theta_4(x)}= m^{\frac{1}{4}} \ 
\text{sn}( \frac{ 2K}{\pi} x), \llabel{jacobi-sn}
\end{equation}
 we find the RHS of Eq.~(\ref{eom-3}) is given by
\begin{eqnarray}
\text{RHS Eq.~(\ref{eom-3})}&=&
\frac{\theta^2_1(\frac{k}{2})}{\theta^2_4(0)} \; \left\{
\frac{\theta^2_1(\frac{\phi}{2})}{\theta^2_4(\frac{\phi}{2})} -
\frac{\theta^2_1(\frac{\phi-k}{2})}{\theta^2_4(\frac{\phi-k}{2})}\right\}
\nn , \\
&=& m^{\frac{1}{2}} \frac{\theta^2_1(\frac{k}{2})}{\theta^2_4(0)}
\ \left\{ \text{sn}^2 \frac{K }{\pi} \ \phi-  \text{sn}^2
\frac{K }{\pi} (\phi- k)\right \}, \llabel{rhs-2}
\end{eqnarray}
where $K \equiv K(m)$ is defined in Eq.~(\ref{elliptic}).
The LHS of Eq.~(\ref{eom-3}) is
\beq
 \frac{1}{2} \ \overline{\omega}_{k}^{2 } \ \frac{d}{d \phi} \  \left[ \frac{\theta'_{4}(\frac{\phi-k}{2})}{\theta_{4}(\frac{\phi-k}{2})} -\frac{\theta'_{4}(\frac{\phi}{2})}{\theta_{4}(\frac{\phi}{2})}\right],\nn
\eeq
and, on using \footnote{This result follows most simply from the Fourier series of both sides.},
\beq
\frac{d}{d z} \left( \frac{\theta'_{4}(z)}{\theta_{4}(z)} \right) = \frac{4 K^{2}}{\pi^{2}} \left[ 1- \frac{E}{K} - m \ \text{sn}^{2} \frac{2 K z}{\pi}\right],
\eeq
we find
\begin{eqnarray}
\text{LHS Eq.~(\ref{eom-3})}&=&  \ \overline{\omega}_{k}^{2 } \ \frac{m K^{2}}{\pi^{2}} \left\{ \text{sn}^2 \frac{K }{\pi} \ \phi-  \text{sn}^2
\frac{K }{\pi} (\phi- k)\right \}.
\end{eqnarray}
Thus Eq.~(\ref{eom-3})
is satisfied provided we assign the frequency as one of two equivalent expressions,
\beq
\overline{\omega}_{k}= \frac{\pi}{m^{\frac{1}{4}}K} \frac{|\theta_{1}(\frac{k}{2})|   }{\theta_{4}(0)} = 
\frac{2   }{\theta'_{1}(0)}
 |\theta_{1}(\frac{k}{2})|, \label{disp-ring}
\eeq
with the help of the following standard
relations~\cite{ww-3} among Jacobi's constants:
\begin{equation}
\theta_1'(0)= \theta_2(0) \theta_3(0)\theta_4(0), \quad
\theta_2(0)\theta_3(0) = \frac{2}{\pi} m^{\frac{1}{4}} \ K.
\llabel{jacobi-constants}
\end{equation}
Equation (\ref{disp-ring}) does not appear to have been published before, but an
equivalent result has been obtained by Sutherland~\cite{bill-unpub} in unpublished notes.

To summarize this section,
the displacement
given by the expression in Eq.~(\ref{sol-periodic}) satisfies the equation of
motion, Eq.~(\ref{eom-3}), for the Toda ring
with the frequency given by Eq.~(\ref{disp-ring}).

\subsection{Connection between the frequency spectrum of the cnoidal waves and
Toda's spectrum.}

Toda's result for the frequency of wave-like solutions of the Toda lattice is
\beq
\overline{\omega}^{{T}}_{k}=  \frac{\pi}{K} \left \{\frac{E}{K}-1 + \frac{1}{\text{sn}^2 \left( \frac{k \ K}{\pi}\right)} \right \}^{-\frac{1}{2}} \label{disp-toda-2}.
\eeq
We give a derivation of this in Appendix \ref{app-toda}.
The difference between our result for the dispersion relation in
Eq.~(\ref{disp-ring}) and Toda's in Eq.~(\ref{disp-toda-2}) is shown
in Fig.~\ref{fig-dispersion}. For small $m$ these are very
close indeed, whereas for $m \to 1$ our solution always has a higher frequency.
The difference between these dispersions also plays a role in determining the
average potential energy as we show below in Eq.~(\ref{pot-average-2}). As
pointed out in Appendix \ref{app-toda}, the dispersion $\omega^{T}_{k}$
comes from a calculation which does not respect the boundary conditions, and
hence the discrepancy with our result is not surprising.
Perhaps what is surprising is that they are so
close for small $k$, and differ significantly only for
$m$ close to unity.

\begin{figure}[!tbh]
\includegraphics[width=\figurewidth]{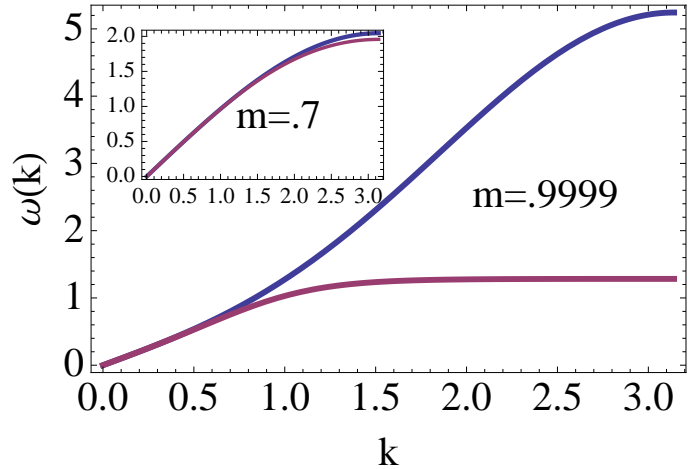}
\caption{The main figure shows the
dispersion relations at $m=0.9999$, which is strongly anharmonic. The
inset is for $m=0.7$, which is only moderately anharmonic.  The upper curve in
both cases is the true spectrum, $\overline{\omega}_{k}$, from Eq.~(\ref{disp-ring})
and the lower one
is $\overline{\omega}^{{T}}_{k}$
from Eq.~(\ref{disp-toda-2}). Except for $m$ very close to unity, the two
spectra are very close.}

\llabel{fig-dispersion}
\end{figure}

In order to appreciate the role of the non linearity,  we may usefully express  the displacement and spectrum as expansions in power of the
elliptic nome parameter $q$ defined in Eqs.~(\ref{q}, \ref{nome-expansion}):
\begin{eqnarray}
\bar{\omega}_k & = &  |2 \sin (\frac{k}{2})| \ [1+ 4 q^2 \sin^{2} (\frac{k}{2}) + 12 q^4 \sin^{2} (\frac{k}{2})  ] +O(q^6), \nn \\
\bar{\omega}^{T}_k& = & |2 \sin (\frac{k}{2})|  \ [1+ 4 q^2 \sin^{2} (\frac{k}{2}) \ \cos(k) +  q^4 \sin^{2} (\frac{k}{2}) \{ 6-5\cos(k)+ 14\cos(2k)- 3\cos(3k)\}  ] +O(q^6) ,  \nn \\
d_k(\phi) & = & 2 q [ \cos (\phi )-\cos (k- \phi ) ] + 
2 q^2 \sin (k) \sin (k- 2 \phi )  \nn \\
&& + \frac{8}{3} q^3 \left[ \cos ^3(\phi )-\cos ^3(k - \phi )\right] + 
q^4 \sin (2 k) \sin [2 (k - 2 \phi )]+ O(q^5) .
\llabel{soln-pert} 
\end{eqnarray}
Here we made use of the expansion of $1/\text{sn}^{2}$ in example 57, page 535 of Ref.~\cite{ww}
to rewrite Eq.~(\ref{disp-toda-2}). Note that the difference between the first two lines appears at $O(q^{4})$, and  is  amplified for large $k$.

\begin{figure}[!tbh]
\includegraphics[width=3in]{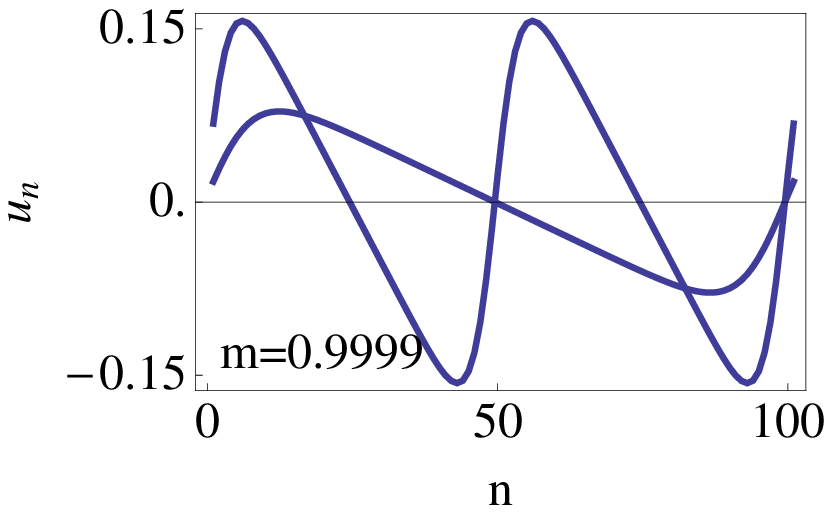}
\includegraphics[width=3in]{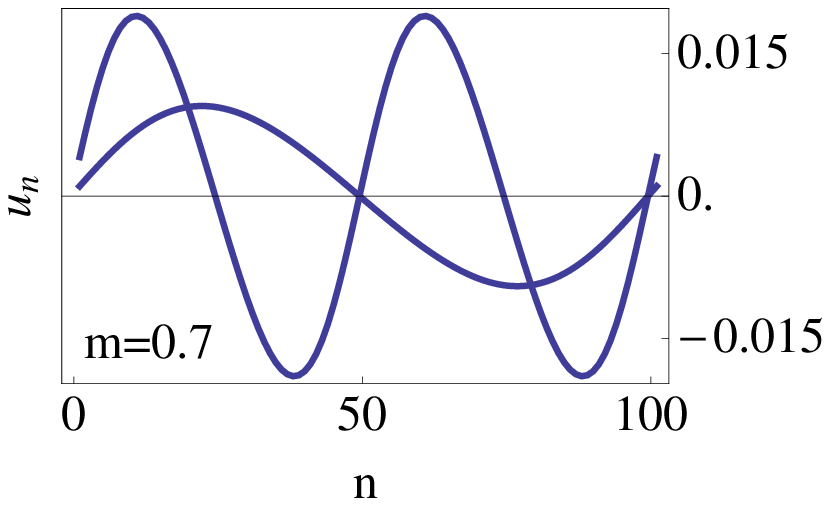}
\caption{The displacements at a fixed time for different sites obtained from
Eq.~(\ref{sol-periodic}). The data is for 
$m=0.9999$, which is strongly anharmonic, 
and $m=0.7$ where the anharmonicity is weaker.
Here $b=1$ and the ring is of length $N=100$. In each case,
results are shown for the two
smallest values of $k$, see Eq.~(\ref{kvalues}).
}
\llabel{fig-displacements}
\end{figure}

In Fig.~(\ref{fig-displacements}) we plot the displacements at different
sites,
obtained from Eq.~(\ref{sol-periodic}),
for a ring of length $N=100$ at two different values of the parameter $m$ with
the lowest two $k$-values, see Eq.~(\ref{kvalues}). Notice the asymmetric shape of the
displacements and their relatively broad structure.

To further visualize the periodic solution, we first note that
Eq.~(\ref{sol-periodic}) is closely related to the singly periodic Jacobi zeta
function $Z(u)=Z(u+2K)$ via the relation $\frac{d}{d \phi} d_{k}(\phi) =
\frac{K}{\pi} [Z(\frac{K}{\pi} (\phi-k) )-Z(\frac{K}{\pi} \phi )]. $ The
Jacobi zeta function has a formal Fourier series expansion, as well as one in
terms of tanh.
 These are 
\begin{align}  
Z(u)&= \frac{\pi}{2K} \ \frac{\theta'_4 (\frac{\pi u}{2 K}) }{\theta_4
(\frac{\pi u}{2 K})}, \label{jacobi-zeta}\\
&= \frac{2 \pi}{K} \sum_{n=1}^{\infty}
\frac{q^n}{(1-q^{2n})} \ \sin \frac{\pi n u}{K} , \;\; \mbox{valid for} \;\; \Im m (u) \leq K';
\label{jacobi-zeta-1}
\\
&= -\frac{\pi}{2 K K'} u + \frac{\pi}{2 K'}   \sum_{\nu=-\infty}^{\infty}
\tanh \{ \frac{\pi}{ 2 K'} ( u - 2 K \nu) \}, \llabel{zeta}
\end{align}
where the first (second)  series expansion is particularly useful in the
limit $m \rightarrow 0$ ($m \rightarrow 1$)\cite{fnote}.
The solution of the Toda lattice on a ring, i.e.~the ``cnoidal wave'',
corresponds to
\begin{align}
d_k(\phi)&= \frac{K}{\pi} \int_\phi^{\phi-k} \ d\phi' \ Z(\frac{K}{\pi} \phi')
\, ,
\nn \\
&= - 4 \sum_{n=1}^\infty \frac{1}{n} \ 
\frac{q^n}{1 - q^{2n}}\ \sin n (\phi-\frac{k}{2}) \ \sin n \frac{k}{2} , \nn \\
&= \frac{K }{4 \pi  K'} ( 2 \phi- k) k + \sum_{\nu= -\infty}^{\infty} \log\left[ 
\frac{\cosh \frac{K}{2 K'} \{ \phi - k-  2 \pi \nu \}}{\cosh
\frac{K}{2  K'} \{ \phi- 2 \pi \nu \}}
\right]. 
\llabel{toda-solution}
\end{align}
The last line follows from  using the  Poisson summation  formula to rewrite the second line\cite{fnote}.

Let us  comment on the limit when the elliptic parameter $m$ tends to 0, where the second line in Eq.~(\ref{toda-solution}) is useful.  It is easily seen that we obtain
the harmonic excitations in this limit.  Using the standard expansion of the various objects in $m$, the displacement is given by
$u_j= \frac{m}{8 b} (\cos(\phi_j)- \cos(\phi_j- k) )$, and the prefactor of $m$ makes these small amplitude oscillations. 
The spectrum is also that of the harmonic limit,
since
Eq.~(\ref{disp-ring}) becomes $\bar{\omega}_k =  2 |\sin \frac{k}{2} |$ and the
phase factor in Eq.~(\ref{const-profile-1}) is
given by $\phi_j= k ( j - c_k t ) $ with
velocity $ c_k= \sqrt{a \ b/ M}= c_0 
$, as expected from Eq.~(\ref{harmonic-1}).  

In the other extreme limit of the elliptic parameter, $m \rightarrow 1$, the elliptic functions degenerate into hyperbolic functions and the cnoidal wave is regarded as a train of solitons, so that  the periodicity of the
 displacements around the ring is  unimportant. 
This is illustrated in Fig.~(\ref{fig-expdus}),
where we plot, for a typical case, the displacements $u_{j}$, their nearest neighbor differences $u_{j} -u_{j+1}$, and the exponential of the latter, i.e. $\exp{\{u_{j} -u_{j+1}\}}$.
\begin{figure}[!tbh]
\includegraphics[width=3in]{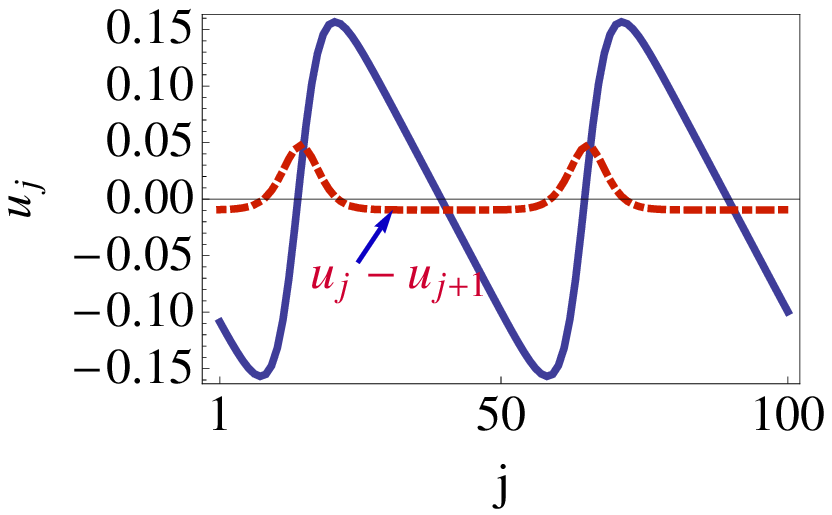}
\includegraphics[width=3in]{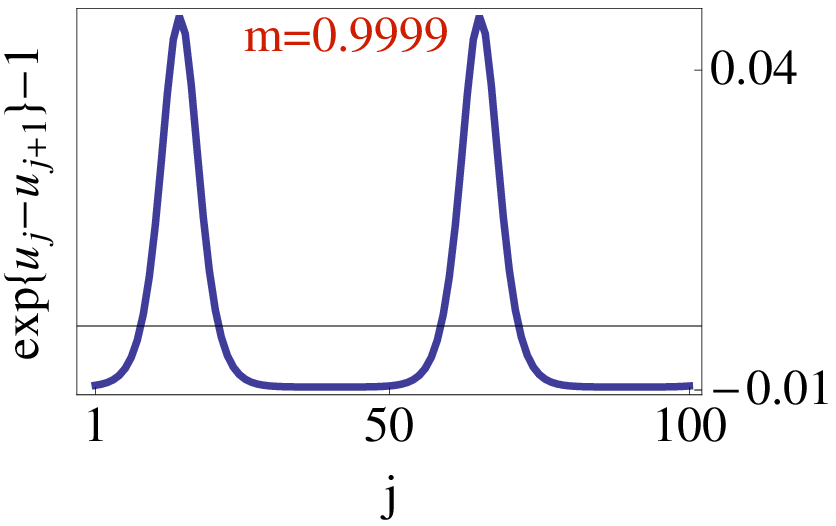}
\caption{
The isolation of a soliton from the cnoidal wave is illustrated here. We plot
various objects for $m=0.9999$, $b=1$ and a ring length $N=100$ with   $k = 4
\pi/N$.    In the left panel we display  the cnoidal wave displacements
$u_{j}$, and  the difference between successive displacements $u_{j}-u_{j+1}$.
In the right panel, we display the exponential of the displacement difference.
The latter shows two Toda solitons, with a clear separation $\sim N/2$ and a
width that is several lattice constants, but much less than the separation.
Note that  while the displacements
appear delocalized and wave like, the solitons are quite localized and hence
particle like.
}
\llabel{fig-expdus}
\end{figure}
 
Let us now extract a single isolated soliton from the solution. We start
with Eq.~(\ref{pot-11}) for the exponential of the displacement difference in
terms of $\text{sn}^2$ and using $\text{dn}^2(u)=1-m \ \text{sn}^2(u)$,
Eq.~(\ref{zeta}) and the relation $\frac{d}{d u} Z(u)= \text{dn}^2(u)- E/K$, we
find  a useful and formally exact series
representation\cite{toda,toda-book,toda_2}
\beq
e^{ b  u_j - b  u_{j+1}} = A + \overline{\omega}^2_k \ (\frac{K}{2 K'})^2 \  \sum_{\nu= - \infty}^{\infty} \text{sech}^2  a \frac{K}{\pi}(\phi_{j}(t)-  2 \pi \nu),\label{disp-series}
\eeq
where $a= \frac{\pi}{2 K'}$ and $$A= (\frac{\overline{\omega}_{k} \ K}{\pi})^2
\{ \frac{\text{cn}^2(\frac{k K}{\pi})} {\text{sn}^2(\frac{k K}{\pi})}+
\frac{E-a}{K} \}. $$ The periodicity in the phase angle $\phi\to \phi+ 2 \pi
\nu$ is manifest in this way of writing the displacement difference.

Let us
focus on $m \to 1$ so that we may set $K'\to \pi/2$ and $a \to 1$. With
$\phi_{j}(t)= k j - c_{0 } \ \overline{\omega}_{k} t$, we observe that the separation
$\Delta j$  between peaks  of the exponential of displacement difference
Eq.~(\ref{disp-series}), is $2 \pi/k$.  We shall see in
Eq.~(\ref{disp-series-3}) below that the width of
these peaks is given by $\kappa^{-1}$ where
\begin{equation}
\kappa = {1 \over \pi} \, k \, K,
\label{kappa}
\end{equation}
and so the requirement that the separation between the peaks is much greater
their separation is
\begin{equation}
K \gg 1, \ \ \text{or equivalently }\ 1-m \ll 1 \, .
\label{Kgg1}
\end{equation}
It is clearly necessary that the width of the peak is at least several lattice
spacings, and so we also need the condition $\kappa \lesssim 1$. Combining with
Eqs.~(\ref{kappa}) and (\ref{Kgg1}) the condition for the oscillations to be
described by well separated solitons is
\begin{equation}
k \lesssim  K^{-1} \ll 1 ,
\label{kK1}
\end{equation}
which, as shown in Eq.~(\ref{Kgg1}), implies that $m$ is very close to 1, i.e.~the
system is in the extreme anharmonic limit.

In the proximity of a peak, we drop the sum over $\nu$, set
$A \to 1$~\footnote{This follows since in the limit $m\to 1$, $ k \to 0 $, we have the behavior  $\text{sn}(kK/\pi)\to \sinh(k K/\pi)$,  $\text{cn}(kK/\pi)\to 1$,  and $E\to 1, \ a \to 1$, and we take into account Eq.~(\ref{energy-soliton}). }, and write
\beq
e^{b(u_j - u_{j+1})} = 1 + \overline{\omega}^2_k \ \left(\frac{K}{\pi}\right)^2
\  {\rm sech}^2 \left[ \left(\frac{K}{\pi}\right)
\left(k j - c_{0}\ \overline{\omega}_{k}t \right) \right].
\label{disp-series-2}
\eeq
From Eq.~(\ref{disp-ring}), $\theta_{1}(k/2)$ determines the dispersion,  and in  the expression Eq.~(\ref{theta1-jacobi-transform}),
the first few terms in an expansion for positive $k$ read as
\begin{equation}
\bar{\omega}_k = \frac{2 K'}{K} e^{{- k^{2} K/K'}}\, \left[  \sinh \frac{k K}{2 K'} - e^{- 2 \pi \frac{K}{K'}} \sinh \frac{3 k K}{2 K'} \right] .
\end{equation}
As $m \to 1$,  $K$ is large and $K'\to \pi/2$, and in the regime $k K \lesssim 1$ which is well satisfied here,
see Eq.~(\ref{kK1}),
we may further approximate this by writing 
\begin{equation}
\bar{\omega}_k = \frac{\pi}{K}\,  \sinh\left(\frac{k K}{\pi}\right) . \llabel{energy-soliton}
\end{equation}
The same answer is also found from Toda's relation Eq.~(\ref{disp-toda-2}), since 
$k$ is small enough in this regime that
the distinction between the two dispersions is negligible.
In terms of the parameter $\kappa$ defined in
Eq.~(\ref{kappa})
above, and calling the soliton
velocity 
\begin{equation}
c_{\kappa}= c_{0} \frac{\sinh \kappa}{\kappa} ,
\label{ckappa}
\end{equation}
where $c_{0}= \sqrt{\frac{a b }{M}}$, we find
\beq
e^{b(u_j - u_{j+1})} - 1 = \frac{ \sinh^2 \kappa}{  \cosh^2
\left[ \kappa ({j} - c_{\kappa} t)  \right]}.
\label{disp-series-3}
\eeq
Equation (\ref{disp-series-3})
is the profile of the famous Toda
soliton, two of which are seen in the right panel of Fig.~\ref{fig-expdus}.
We see that the soliton parameter $\kappa$ controls the amplitude
of the solitons (through the prefactor in the amplitudes $\sinh^2( \kappa)$),
the velocity $c_\kappa$ and also the length scale of spatial  variation
$\frac{1}{\kappa}$.  Since $ \frac{\sinh \kappa}{\kappa}\geq 1$, the velocity of the soliton
$ c_{\kappa}$ in Eq.~(\ref{ckappa})
is always greater than the sound velocity 
$c_{0}$.

For completeness  we note the displacement for an isolated soliton,
by using the same limits as above in Eqs.~(\ref{Kgg1}, \ref{kK1}). 
The displacement $u_{j}$ and its time derivative can be written   from
Eq.~(\ref{toda-solution}), where we retain only $\nu=0$ (since $K\gg1$) and
write
\begin{eqnarray}
u_j  & \sim & \frac{1}{b} \    \log\left[ 
\frac{\cosh  \kappa (j-1 -c_k t)  }{\cosh  \kappa (j -c_k t) 
 } \right] ,
 \label{uj}\\
 \dot{u}_j & \sim & \frac{c_0 \sinh\kappa }{b}  \left[ \tanh \kappa(j - c_k t)
- \tanh \kappa(j -1- c_k t) \right] .
\label{ujdot}
\end{eqnarray}
It is now straightforward to see that the equation of motion Eq.~(\ref{eom-2})
is satisfied by the single soliton solution given in
Eqs.~(\ref{disp-series-3})--(\ref{ujdot}). Toda notes that if we ignore the periodic
boundary conditions, and imagine Eqs.~(\ref{disp-series-3})--(\ref{ujdot}) to be
extended for all $-\infty \leq j \leq \infty$, then integrating Eq.~(\ref{uj})
gives
$u_{\infty} - u_{-\infty}= \frac{ 2\kappa}{b}$, so there is
a net compression near the soliton. Hence the soliton can be regarded as a
local compression propagating with speed $c_\kappa$ given by
Eq.~(\ref{ckappa}).

The momentum  $M \dot{u}_{j}$ vanishes at $j\to \pm
\infty$, as does the potential energy term Eq.~(\ref{disp-series-3}).
Therefore the energy of a soliton, obtained by ignoring periodic
boundary conditions and opening the ring into an infinite chain, is finite. 
Substituting for $\dot{u}_{j}$ and the displacement difference from
Eqs.~(\ref{uj})--(\ref{ujdot}) into the Hamiltonian Eq.~(\ref{toda-H}),
and with $p_{j}= M \dot{u}_{j}$,  we obtain this energy to be
$\varepsilon_{\text{Soliton}}=\frac{2 a}{b} (\sinh \kappa \cosh \kappa - \kappa)$.

With the Fourier expansion in Eq.~(\ref{toda-solution}) one can give an alternate
derivation of the dispersion relation of the Toda lattice with periodic boundary
conditions in Eq.~(\ref{disp-ring}).  This is done in
Appendix~\ref{alt-deriv}.

\section{Energy of cnoidal waves}

\llabel{energy}
We next turn calculate the total energy of the Toda
ring with a
cnoidal wave.  As a prelude, let us recapitulate the results of the
trivial harmonic lattice, with the Hamiltonian in Eq.~(\ref{toda-H-harmonic})
truncated at the quadratic level.  If we assume a phonon displacement
$u_n= u_0 \sin( k n - \omega_k t)$, then $\omega_k =
\sqrt{\frac{\kappa}{M}} | 2 \sin \frac{k}{2} |$. The cycle average of
the total kinetic energy and potential energy at wavevector $k$
are easily found to be
\begin{eqnarray}
\overline{\text{KE}} &=&\overline{\text{PE}} = N u_0^2 \kappa \sin^2
\frac{k}{2} , \nn \\
\rho & \equiv & \frac{\overline{\text{KE}}}{\overline{u_n^2}} = 
\frac{1}{2} N M \omega_k^2 = \frac{1}{2} N a b\, | 2 \sin \frac{k}{2} |^2.
\llabel{harmonic}
\end{eqnarray}

We next calculate the kinetic energy of the Toda ring using the cnoidal
wave solution Eq.~(\ref{toda-solution}) as a Fourier series.  Let us start with the kinetic energy
expression:
\begin{eqnarray}
\text{KE}&=& \frac{M}{2} \sum_j (\dot{u}_j)^2\nn \\
&=& \frac{M\omega_k^2}{2 b^2} \sum_j \left\{ d'_k(\phi_j(t)) \right\}^2 \nn \\
\overline{\text{KE}}&=& 4\ N \left( \frac{a}{b} \ \bar{\omega}_k^2 \right)
\sum_{n=1}^\infty \frac{q^{2 n}}{(1- q^{2n})^2} \ \sin^2 n \frac{k}{2}.
\llabel{kinetic}
\end{eqnarray}
Here we used Eq.~(\ref{const-profile-1}), and the last line is obtained by
taking the time average over a cycle, with the displacement from the  series in
Eq.~(\ref{toda-solution}). 

The potential energy of the Toda ring,
averaged over a cycle, is calculated in Appendix \ref{potential-energy}, where we show that
\beq
\overline{  \text{PE}}= N \frac{a}{b} \left[  \left(\frac{\overline{\omega}_{k}}{\overline{\omega}^{T}_{k}}\right)^{2}-1\right].\label{pot-average-2}
\eeq
We expand Eq.~(\ref{pot-average-2}), as in Eq.~(\ref{soln-pert}) in terms of the nome ``q'' to lowest order and find
\beq
\overline{ \text{PE}}= N \frac{a}{b} \left[ 16  q^{2 }\sin^{4}(\frac{k}{2}) (1+ 6 q^{2}) +O(q^{6}) \right]. \label{pot-pert}
\eeq
We  now compute the ratio of kinetic to potential energies, and express it as a series in the nome:
\beq
\frac{\overline{\text{KE}}}{\overline{\text{PE}}}= 1+ 4 q^{2}\sin^{2}(\frac{k}{2})+O(q^{4}),
\eeq
which is unity for small anharmonicity, as expected, and increases above unity
for greater anharmonicity. 

In fact, we shall mainly compute the Fourier
series for the mean square displacement
\begin{eqnarray}
u_j & = & \frac{1}{b} d_k(\phi_j(t)) , \nn \\
\overline{u^2_j} &=&  \frac{8}{b^2} \sum_{n=1}^\infty \frac{1}{n^2} \ 
\frac{q^{2n}}{(1 - q^{2n})^2}\  \ \sin^2 n \frac{k}{2} \nn \\
&=& \frac{2}{b^2} \int_0^{\frac{k K}{2 \pi}} dy \ y  \ \xi(\frac{k K}{2 \pi}-y), 
\llabel{fourier-amplitude}
\end{eqnarray}
where $\xi(u)= \int_0^{2K} \frac{d v}{2K} Z(v)Z(u+v)$
\footnote{Here $k$ stands for the wave vector
and should not be confused with the 
elliptic function modulus used in standard books~\cite{ww}. 
As mentioned earlier, we will  use the elliptic 
parameter ``m'' consistently
in this work, and not  the elliptic modulus``k''. }.
Combining Eqs.~(\ref{fourier-amplitude}) and Eq.~(\ref{kinetic}), we find
the total energy to mean square amplitude ratio
\begin{equation}
\rho \equiv \ \frac{\overline{\text{KE}}}{\overline{u^2_j}}=
\frac{1}{2}\ ( N M \overline{\omega}_k^2) \frac{\sum_{n=1}^\infty
\frac{q^{2 n}}{(1- q^{2n})^2} \ \sin^2 n \frac{k}{2}}
{\sum_{n=1}^\infty \frac{1}{n^2} \ \frac{q^{2n}}{(1 - q^{2n})^2}\  \ 
\sin^2 n \frac{k}{2}}, 
\llabel{ratio}
\end{equation}
with Eq.~(\ref{disp-ring}) defining the spectrum $\overline{\omega}_k$. 

We can express the average kinetic energy in a more compact form using
the expression in Eq.~(\ref{pot-1}) for the displacements. Let us write
\begin{eqnarray}
\overline{\text{KE}}&=&- \frac{M}{2} \sum_j \overline{ \ddot{u}_j \ u_j}\nn \\
&=& \frac{N  a}{2 b} \ \frac{\omega_k}{2 \pi} \  \int_0^{\frac{2 \pi}{\omega_k}}
\  \ d t \ \{d_k(\phi- k)- d_k(\phi)\} e^{\{d_k(\phi- k)- d_k(\phi)\}} \nn \\
&=& \frac{N  a}{2 b} \ \alpha_k \  \ \ \int_0^{2 K} \ \frac{du}{2K} \ 
\left (  \text{dn}^2 u -  \frac{E}{K} \right ) \log \left[
1+ \alpha_k \  ( \text{dn}^2 u -  \frac{E}{K} )   \right ] ,
\nn \\ 
\llabel{kinetic-2}
\end{eqnarray}
where we have set $ \alpha_k= \bar{\omega}_k^2 \frac{K^2}{\pi^2}$ and  used
Eq.~(\ref{pot-11}).
This expression is particularly useful if we want long wave length
results since we can expand in $\alpha_k$ and find the exact elliptic
parameter (i.e.~$m$) dependent coefficients. The leading term for long
wavelengths (up to and including  $O(k^4)$) is
\begin{eqnarray}
\overline{\text{KE}} & = & \frac{N  a}{2 b}  \alpha_k^2 \   
\left[ \ \int_0^{K} \ \frac{du}{K} \ dn^4 u - \frac{E^2}{K^2} \right]+ O(k^4), \nn \\
&=& \frac{N  a}{2 b}  \alpha_k^2 \left[ 
\frac{2}{3}(2-m) \frac{E}{K}-\frac{1}{3}(1-m)
- (\frac{E}{K})^2 \right]  + O(k^4), \nn \\
& \sim &
\left\{
\begin{array}{ll}
\displaystyle \frac{N  a}{2 b}  \alpha_k^2 
\left[ \frac{m^2}{8}  -\frac{m^4}{1024}\right],  &
(m \rightarrow 0), \\
\displaystyle \frac{N  a}{2 b}  \alpha_k^2 
\left[ \frac{4}{3 \log\frac{16}{1-m}}\right], &
(m \rightarrow 1) .
\end{array}
\right.
\end{eqnarray}

We can also  extract the long wave length behavior of
Eq.~(\ref{fourier-amplitude}) using~\cite{as-2} as
\begin{eqnarray}
\overline{u^2_j} & = &  \frac{2}{b^2} k^2 \ \sum_{n=1}^\infty 
\ \frac{q^{2n}}{(1 - q^{2n})^2} + O(k^4), \nn \\
&=& \frac{2}{b^2} k^2 \left[ \frac{1}{24}+ \frac{1}{6}(2-m) 
\frac{K^2}{\pi^2} + \frac{E K}{2 \pi^2} \right]\nn \\
& \sim &
\left\{
\begin{array}{ll}
\displaystyle \frac{2}{b^2} k^2 
\left[\frac{m^2}{256} \right] + O(k^4),  &
(m \rightarrow 0), \\
\displaystyle \frac{2}{b^2} k^2 
\left[\frac{(\log(1-m))^2}{24 \pi^2} \right], &
(m \rightarrow 1).
\end{array}
\right.
\end{eqnarray}
We finally put together the results for the kinetic energy and the mean square
amplitude to form the ratio
\begin{subequations}
\llabel{ratios-2}
\begin{align}
\label{smallm}
\rho & = \frac{1}{2} N (a b) k^2
\left[ 1+ \frac{3}{256} (m^2 + m^3) + O(m^4)\right], \ \qquad
(m\rightarrow 0),
 \\
\rho & =  N (a b) k^2 \ 
\frac{\log(16)^3}{ 2 \pi^2} \ \log\frac{1}{1-m}, \qquad\qquad\qquad\qquad
(m\rightarrow 1). 
\end{align}
\end{subequations}
We see that the small m limit of Eq.~(\ref{smallm}) gives the long wavelength limit
of the harmonic lattice
result given in Eq.~(\ref{harmonic}).
In the opposite, strongly anharmonic, limit the
kinetic energy grows logarithmically as $m \rightarrow 1$. 

\section{Conserved Currents of the Toda Ring}

\llabel{conserved}
We summarize in this section the construction of the conservation laws
of the Toda lattice, and write out explicitly the first few conservation
laws. For brevity we set $m=a=b=1$. The work of Henon~\cite{henon}
and Flaschka~\cite{flaschka} gives
us a construction of the conservation laws starting with the Lax
equation $\dot{L}=[L,A]$, where $L$ and $A$ are $N \times N$ matrices
with entries
\begin{eqnarray}
L_{n,m} &= & \delta_{n,m} p_n + \chi_{n,m} , \nn \\
\chi_{n,m} & = & i \{ \delta_{m,n+1} e^{-\frac{1}{2} (u_m-u_n)}- 
\delta_{m,n-1} e^{-\frac{1}{2} (u_n-u_m) } \} , \nn \\
A_{n,m} &= & -i/2 \ \{ \delta_{m,n+1} e^{-\frac{1}{2} (u_m-u_n)} +
\delta_{m,n-1} e^{-\frac{1}{2} (u_n-u_m) } \} , \llabel{lax-1}
\end{eqnarray}
so that the Lax equation leads  to the original equation of motion,
Eq.~(\ref{eom-1}). Further, we can construct  $N$ independent conservation laws by
taking the trace of the first $N$ powers of the Lax matrix. With the definition 
\begin{equation}
J_n=
\frac{1}{n!} \mathrm{Tr}\, L^n;  \;\;\; 1 \leq n \leq N \label{lax-currents}
\end{equation}
it has been shown that these are in mutual involution~\cite{henon,flaschka,toda}, i.e. 
their Poisson brackets are all
zero, $\{J_{n},J_{m} \}=0$, and thus
they form a complete set of independent conservation laws.   We list the first few conservation laws:
\begin{subequations}
\begin{eqnarray}
J_1&=& p_\text{total}\, ,  \\
J_2 &=& H +\text{const.} \, , \\
J_3 &= &\frac{1}{6} \sum_n p_n^3 + \frac{1}{2}
\sum_n p_n \ \{e^{{ (u_{n}-u_{n+1})}}+ e^{{ (u_{n-1}-u_{n})}} \} \, ,
\llabel{J3}\\
J_4 &=&  \frac{1}{24} \sum_n p_n^4 + \frac{1}{6} \sum_n p_n^2 
\{e^{{ (u_{n}-u_{n+1})}}+ e^{{ (u_{n-1}-u_{n})}} \}+ \frac{1}{6} \sum_n p_{n+1} \ p_n 
\ e^{{ (u_{n}-u_{n+1})}} \nn \\
&&+ \frac{1}{6} \sum_n  \ e^{{ (u_{n}-u_{n+2})}} + 
\frac{1}{12} \sum_n  \ e^{{ 2 (u_{n}-u_{n+1})}}\, . 
\end{eqnarray}
\llabel{cons-laws}
\end{subequations}
We make extensive use of these conservation laws in later sections.

\section{Persistence of Energy
Current and Its relationship to Conserved Currents}
\llabel{persistence}

In this section we study the decay and persistence of the energy current
in the Toda lattice with periodic boundary conditions.
The energy current
is obtained
from the energy density conservation law $\dot{H}(x,t)+ \partial_x
J_E(x,t)=0$, with a suitable discretization of the spatial derivative.
To obtain the energy  current, we write $H=\sum_j p_j^2/(2 M)+ \frac{1}{2}
\sum_{i,j} V_{i,j}$, where
$V_{i,j}= V(u_i-u_j)= \frac{a}{b} e^{- b (u_i - u_j) \eta(i,j)}$ and
$\eta(i,j)=(j-i) \delta_{1,|i-j|}$. The constant and linear term in
Eq.~(\ref{toda-H}) can be omitted safely since they do not change the current.
Thus the force on atom $i$ due to atom $j$ is $$F_{i,j}= - \partial_{u_i}
V_{i,j} = a \ \eta(i,j) e^{-b (u_i-u_j) \eta(i,j)}.$$ 
We note the symmetry $V_{i,j}=V_{j,i}$ and $F_{i,j}=-F_{j,i}$. Thus we write
$H=\sum_i H_i$ with $H_i= p_i^2/(2 M) + \frac{1}{2} \ \sum_j V_{ij}$, and
therefore denoting $v_i= p_i/M$, 
the rate of change of local energy is given by
\barray
\dot{H}_i & = &  v_i \ \dot{p}_i - \frac{1}{2} \sum_j F_{i,j}\, (v_i-v_j) =
\frac{1}{2} \sum_j F_{i,j}\,(v_i+v_j), \nn \\
&=& \frac{1}{2} F_{i,i+1}\,(v_i+v_{i+1})-F_{i-1,i}\,(v_i+v_{i-1}) , \nn \\
&=& - J_E(i+1) + J_E(i) .
\earray
Therefore we may write alternate  expressions for the energy current that are equivalent:
\begin{align}
J_E& =   \frac{1}{2} \sum_i F_{i-1,i}\, (v_i+v_{i-1}), \;\;\text{or} 
& J_E & =   \frac{1}{2} \sum_i v_i\,( F_{i-1,i}+ F_{i,i+1}),\llabel{energy-current}
\end{align}
\\
with $F_{i-1,i} = -a e^{-b(u_i-u_{i-1})}$. We use the second form of the
current above, and  will set set $M=a=b=1$ in the Hamiltonian
Eq.~(\ref{toda-H}) in the sequel.

We compute the correlation function
\begin{equation}
C_{J_E}(t)= \langle J_E(t) J_E(0) \rangle,
\llabel{cf-1}
\end{equation}
where the average is taken in the canonical ensemble. This function does
not decay to zero at long times but rather reaches a finite value. This
is the phenomenon of temporal persistence, and is related to the
integrability of the underlying model. The implication of this non
vanishing of $C(t\rightarrow \infty)$ is that the Fourier transform of
$C(t)$, namely the Kubo thermal conductivity,  contains a Dirac delta
function at zero frequency $\delta(\omega)$. Thus the thermal
conductivity has a Drude term in common with many other integrable models studied in recent years.

Inspired by Mazur~\cite{mazur}, we assume that the time integrated
energy current can be written as a linear combination of all the conserved
quantities of the model $I_{n}$ 
\begin{equation}
\lim_{t\to\infty} \frac{1}{t} \int_0^t J_E(t') \, dt' = 
\sum_n a_n I_n \, ,
\llabel{expand-curr}
\end{equation}
for \textit{fixed} coefficients $a_n$. Both sides of this equation depend on
the initial state of the system. Again following Mazur, we argue
that it holds, with the same
$a_n$, for almost all
initial states on a constant energy surface. Here we will test
Eq.~(\ref{expand-curr}) for different
initial states from a \textit{canonical distribution}. 
The set of constants of motion $I_{n}$ are not spelled out in detail in earlier work, 
although one expects that these are {\em functions} of the independent  currents $J_{n}$ in Eq.~(\ref{lax-currents}).

Using Eq.~(\ref{expand-curr}), and a few chosen conserved currents, one can obtain a
{\em  lower bound} to the persistent part of the current current
correlations~\cite{mazur}.  The nice thing about this result is that even a
single non trivial conserved 
current could help establish the existence of the temporal
persistence~\cite{quantum-cases}.  While in most applications of this idea, one
has to be content with the result as a \textit{bound}, one would also like to 
test the idea of completeness, i.e.~to see if the bound is
saturated in a case where one has a full knowledge of the conservation laws.  A
natural expectation is that if {\em all }the conserved currents that go into
this expansion are  known, then we should be able to get the exact value of
the persistent part of the correlations.  This is the operator analog of
expanding vectors in a complete basis. Mazur gives the
example of the XY model of magnetism in one
dimension~\cite{niemeijer}, where all the
higher conservation laws  are known in the quantum theory, since 
the problem is reducible to
free fermions. In that case, he points out that exact persistent part of
correlations can be obtained from a knowledge of these conservation laws. 
Initially, we will test the hypothesis that  the set of conservation laws coincides with
the constants of motion found in Eq.~(\ref{lax-currents}), i.e.~we begin by
assuming that  $I_{n}=J_{n}$. Later, we  will discover that one needs to
expand the set $I_{n}$ to include further terms, such as bilinears in the $J_{n}$.
We begin by determining 
the $a_n$ in terms of thermally averaged correlation functions.
Multiplying Eq.~\eqref{expand-curr} (with $I_{n}\to J_{n}$)  by one of the conserved currents $J_m$ and
averaging over initial states with a Boltzmann distribution gives
\begin{equation}
\langle J_E J_m \rangle = \sum_n a_n \langle J_n J_m \rangle\, ,
\llabel{JEJm}
\end{equation}
where we used the fact that $\langle J_E(t) J_m \rangle$ is independent of
$t$.
The $a_n$ are therefore given by
\begin{equation}
a_n = \sum_l \left( C^{-1} \right)_{nl} \langle J_E J_l \rangle \, ,
\llabel{an}
\end{equation}
where $C$ is the matrix of correlations of the conserved currents,
\begin{equation}
C_{nl} = \langle J_n J_l \rangle \, .
\end{equation}
Squaring Eq.~\eqref{expand-curr}, and using Eqs.~\eqref{JEJm} and \eqref{an}, gives
\begin{eqnarray}
\lim_{t\to\infty} C_{J_E}(t) & \equiv & \, \lim_{t\to\infty} \langle J_E(t_0)
J_E(t_0+t)\rangle \nn \\
& = & \lim_{t \to\infty} \left[ \frac{1}{t}
\int_0^t J_E(t') \, dt' \right]^2 \nn \\
& = & \sum_{n,m,l,k} \left(C^{-1} \right)_{nl} \langle J_E J_l \rangle
\left(C^{-1} \right)_{mk} \langle J_E J_k \rangle C_{mn} \nn \\
& = & \sum_{k,l}\langle J_E J_l \rangle  \left(C^{-1} \right)_{lk} 
\langle J_E J_k \rangle \, ,
\llabel{CJE-infinity}
\end{eqnarray}
where we used $\sum_n C_{mn} \,C^{-1}_{nl} = \delta_{m l}$
to obtain the last line.
Noting that $J_E$ is odd under time reversal, only the odd conserved currents,
$n = 1, 3, 5, \cdots, N-1$, in Eq.~\eqref{cons-laws} contribute.

We have investigated the validity of Eq.~(\ref{CJE-infinity}) numerically for
sizes $N = 4, 6$ and 8.
All the numerical results in this paper
are for parameter values $M=a = b = 1$ in the
Toda Hamiltonian, Eq.~\eqref{toda-H}.
We prepare a large number (several thousand) of initial states appropriate to
a temperature $T$ (which we take to be $T=0.5$) using standard
Monte Carlo methods~\cite{newman}. 
Starting from each of these states
we perform molecular dynamics 
using a fourth order symplectic algorithm~~\cite{OMF}, which  
combines high accuracy with long-term stability. We use a time step
$\Delta t$ equal to $0.05$.

\begin{figure}[!tbh]
\includegraphics[width=\figurewidth]{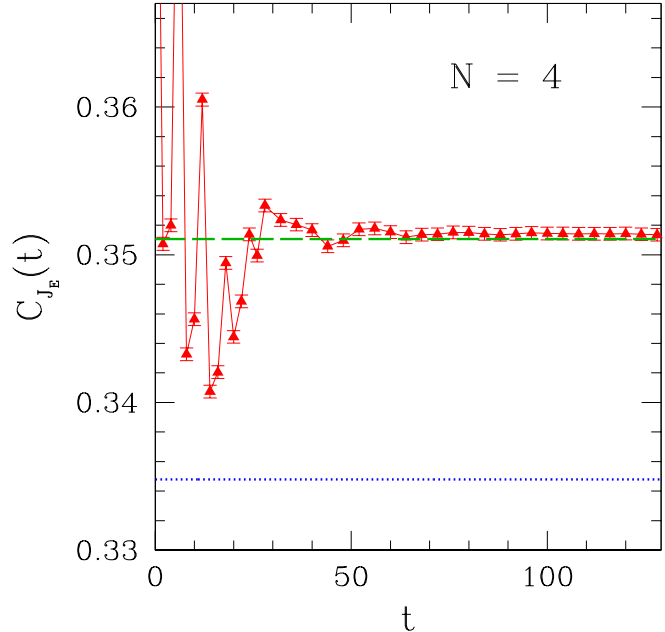}
\caption{The correlation function of the energy current, defined in
Eq.~(\ref{cf-1}), as a function of time for $N=4$ particles. The (blue)
short-dashed
line is the RHS of Eq.~\eqref{CJE-infinity} including all the (odd) Lax currents. 
There is
is a significant discrepancy with the long time limit of $C_{J_E}(t)$.
The (green) long-dashed line is the RHS of Eq.~\eqref{CJE-infinity} including, in
addition, all (odd) pairs of Lax currents.  The agreement with the long time limit
of $C_{J_E}(t)$ is now excellent.
}
\llabel{Ct4_new}
\end{figure}

\begin{figure}[!tbh]
\includegraphics[width=\figurewidth]{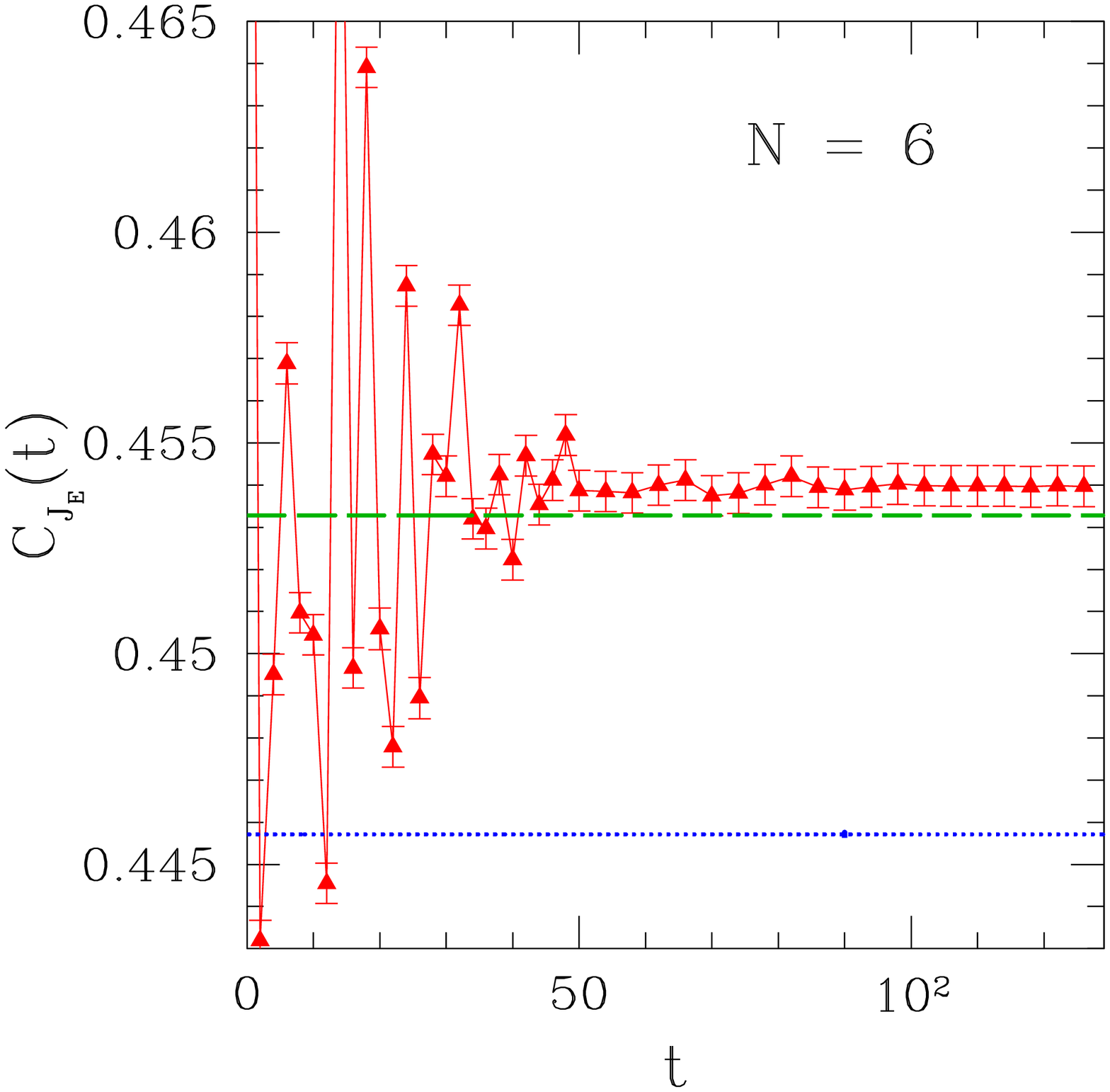}
\caption{The same as Fig.~\ref{Ct4_new} but for $N = 6$.
}
\llabel{Ct6_new}
\end{figure}

\begin{figure}[!tbh]
\includegraphics[width=\figurewidth]{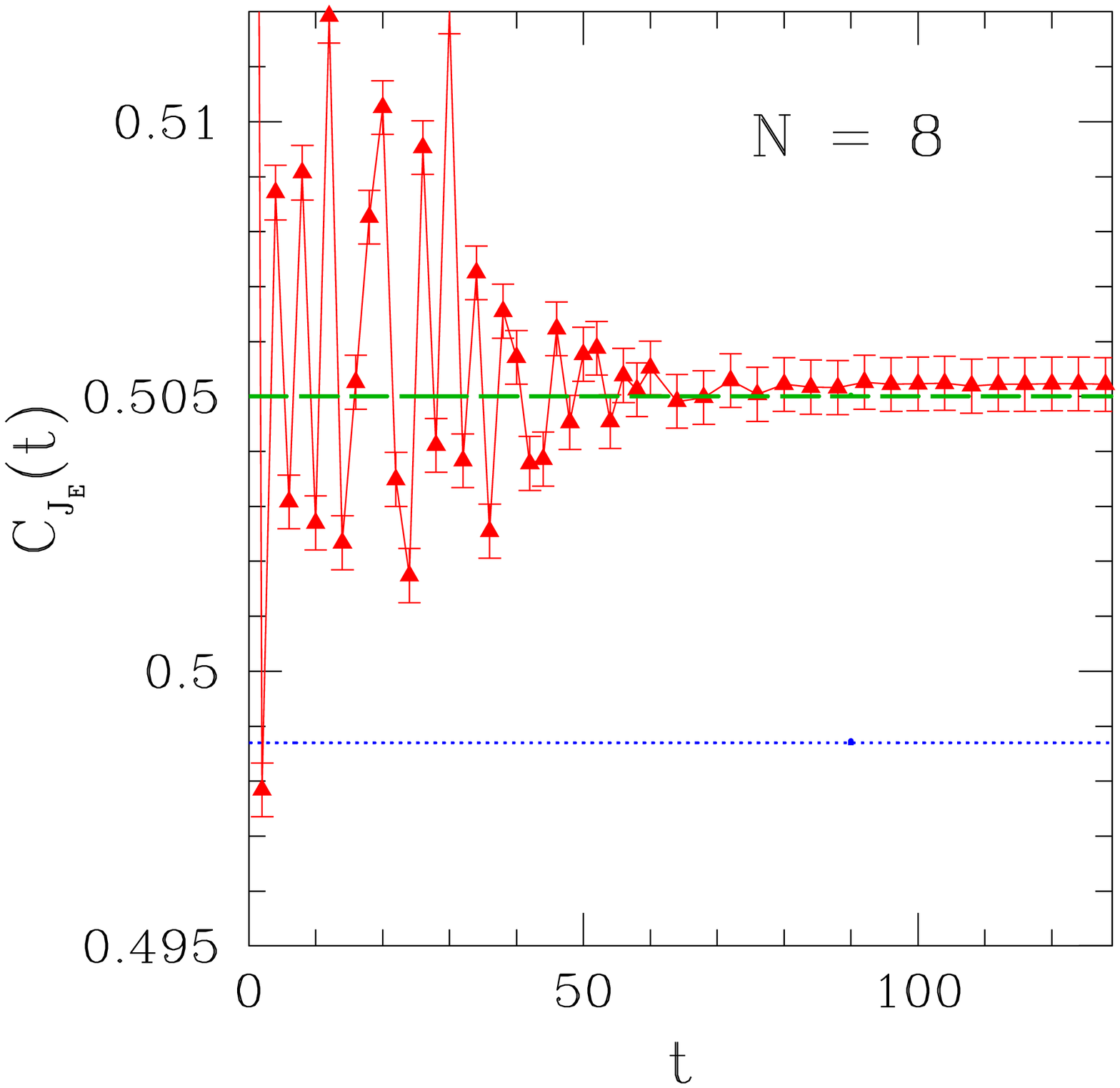}
\caption{The same as Fig.~\ref{Ct4_new} but for $N = 8$.
}
\llabel{Ct8_new}
\end{figure}

Our results are presented in Figs,~\ref{Ct4_new}, 
\ref{Ct6_new} and \ref{Ct8_new} for sizes $N = 4, 6$ and 8 respectively. The
solid line is the data for $C_{J_E}(t)$ and the (blue) dotted line 
is the RHS of Eq.~\eqref{CJE-infinity} including all the (odd) currents. While
the long time limit of $C_{J_E}(t)$ is close to the RHS of
Eq.~\eqref{CJE-infinity} there is a clear discrepancy.

The discrepancy is removed if we note that the Lax currents
$J_n$ in Eq.~(\ref{cons-laws})
are not the only conserved currents but, in addition, \textit{products} of
these currents are conserved.  This corresponds to enlarging the set of
currents $I_{n}$ in Eq.~(\ref{expand-curr}) to include bilinears $ J_n J_l$.

Thus the leading correction to Eq.~\eqref{CJE-infinity} will involve
products of two currents $J_{nl} \equiv J_n J_l$ where one of the indices must
be odd and the other even for $J_{nl}$ to be odd under time reversal. Hence
the total number of conserved ``currents'' to be included in the sums in
Eq.~\eqref{CJE-infinity} is $N/2$ (the odd $J_n$) plus $(N/2)^2$ the
quadratic combinations with odd symmetry. Our results including the quadratic
combinations of currents are shown by the (green) dashed line in
Figs,~\ref{Ct4_new},
\ref{Ct6_new} and \ref{Ct8_new}. The agreement is now excellent.
We thus see that the set of currents that are involved in the expansion must
include not just the Lax currents, but also products of these. 
We see that for the purpose of expanding an operator as in
Eq.~(\ref{expand-curr}), the currents $J_n$ are not a 
\textit{linear} basis, but rather
$J_E$ seems to be \textit{algebraically}
dependent on the $J_n$. Within the numerical
scheme it is hard to determine if we have really saturated the persistent
part by including just first and second powers of the Lax currents, 
but the remainder, if any, must be very small indeed.

We also test Eq.~(\ref{expand-curr}) 
for the case of $N= 4$
by performing a least squares fit to minimize
\begin{equation}
\Delta J_E^2 \equiv
\langle\, \left[ \lim_{t\to\infty} \frac{1 }{N_t} \sum_{m = 1}^{N_t} J_E(t_m)   -
\sum_n a_n J_n \right]^2 \, \rangle
\llabel{lsfit}
\end{equation}
with respect to the $a_n$, where $t_m = m \Delta t$, and
$N_t = t / \Delta t$ is the number of
discrete times in the simulation. Including just the $N/2$ odd currents we
find $\Delta J_E^2 = 0.016886$, which is not extremely small and represents
the difference between the dotted line and the long-time limit of the data in
Figs~\ref{Ct4_new}--\ref{Ct8_new}. However, including $(N/2)^2$ quadratic
combinations of currents that project on to the energy current we get a much
lower value, $\Delta J_E^2 = 0.000220$, as expected from the the good
agreement between the dashed line and the long-time limit of the data in
Figs~\ref{Ct4_new}--\ref{Ct8_new}.
As a consistency check, we compare in Table \ref{tab1} the fit coefficients
obtained by minimizing Eq.~\eqref{lsfit} with those from Eq.~\eqref{an} which
used \textit{equal time} Monte Carlo results. The agreement is good.

\begin{table}[!tb]
\caption{
A comparison, for $N=4$,
between the coefficients $a_n$ in the expansion of the energy current
in terms of the conserved
currents, see Eq.~\eqref{expand-curr}. The ``time-series'' results were
obtained by minimizing $\Delta J_E^2$ in Eq.~\eqref{lsfit}. The ``Monte
Carlo'' results were obtained from Eq.~(\ref{an}).
\label{tab1}
}
\begin{tabular}{|l|r r|r r r r|}
\hline
\hline
 & $a_1$  & $a_3$ & $a_{12}$  & $a_{14}$ &  $a_{32}$  & $a_{34}$  \\
\hline
time series & 0.4965  & $-0.9626$ & 0.1262  & $-0.0591$  &  $0.0015$ & 0.0178 \\
Monte Carlo & 0.4628  & $-0.9412$ & 0.1358  & $-0.0681$  & $-0.0042$ & 0.0227 \\
\hline
\hline
\end{tabular}
\end{table}

 Mazur clarifies that
the above expansion Eq.~(\ref{expand-curr}) should be valid for almost all
initial conditions on the surface of constant energy.  Here we have seen that,
in addition, the expansion works to high numerical accuracy in the
more general case of a canonical distribution.

\section{Perturbed Toda Ring and Decay rates of Persistent Currents}

\llabel{perturb}
We have also studied numerically the decay of persistent currents when a small
perturbation away from integrability is added to the Toda Hamiltonian.
The technique is the same as that described above in Sec.~\ref{persistence}.

We consider two different perturbations:
\begin{equation}
\Delta \mathcal{H} = \left\{
\begin{array}{l}
\displaystyle {w \over 3}\, \sum_{n=1}^N \left(u_{n} - u_{n+1}\right)^3\, , \\
\displaystyle {v \over 4}\, \sum_{n=1}^N \left(u_{n} - u_{n+1}\right)^4 \, .\\
\end{array}
\right.
\llabel{pert}
\end{equation}
The temperature is taken to be $T = 1$, the time step to be $\Delta t =
0.02$, and the lattice has
$N = 64$ particles with periodic boundary
conditions. We have verified that reducing $\Delta t$ or increasing $N$ did
not make a significant difference to the results.

\begin{figure}[!tbh]
\includegraphics[width=\figurewidth]{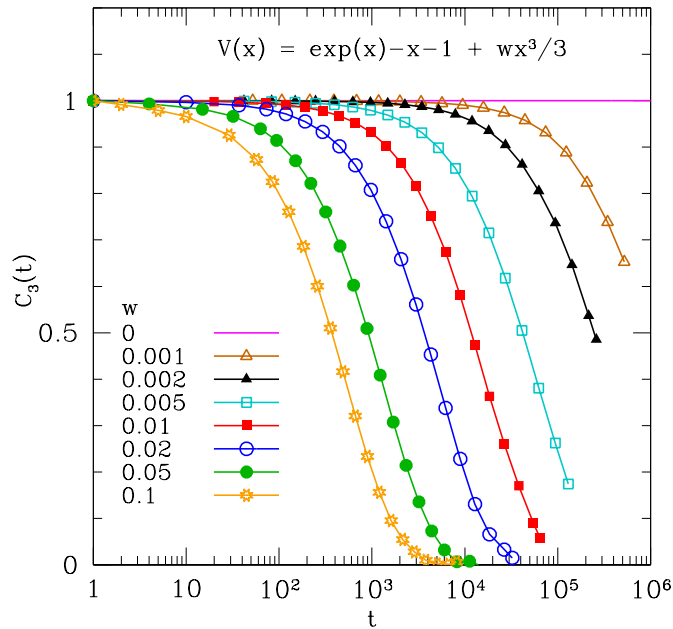}
\caption{The normalized correlation function, defined in Eq.~(\ref{C3}), for the
current $J_3$, for several values of the cubic perturbation $w$
in Eq.~(\ref{pert}) added to the integrable Toda Hamiltonian. The size is
$N = 64$. Including the
perturbation, the system is no longer integrable, so the
correlation function tends to zero on a timescale which diverges as
$w \to 0$.
}
\llabel{J3_tau_several_w}
\end{figure}

\begin{figure}[!tbh]
\includegraphics[width=\figurewidth]{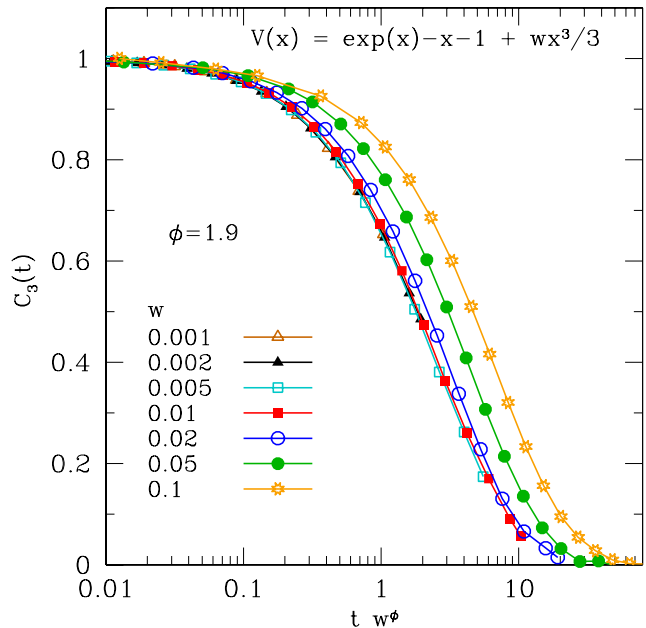}
\caption{A scaling plot,
according to Eq.~(\ref{C3_scale}),
of the data in Fig.~\ref{J3_tau_several_w},
with crossover exponent $\phi = 1.9$ which gives the best
data collapse for small $w$.
}
\label{J3_tau_scale_w_1.9}
\end{figure}

\begin{figure}[!tbh]
\includegraphics[width=\figurewidth]{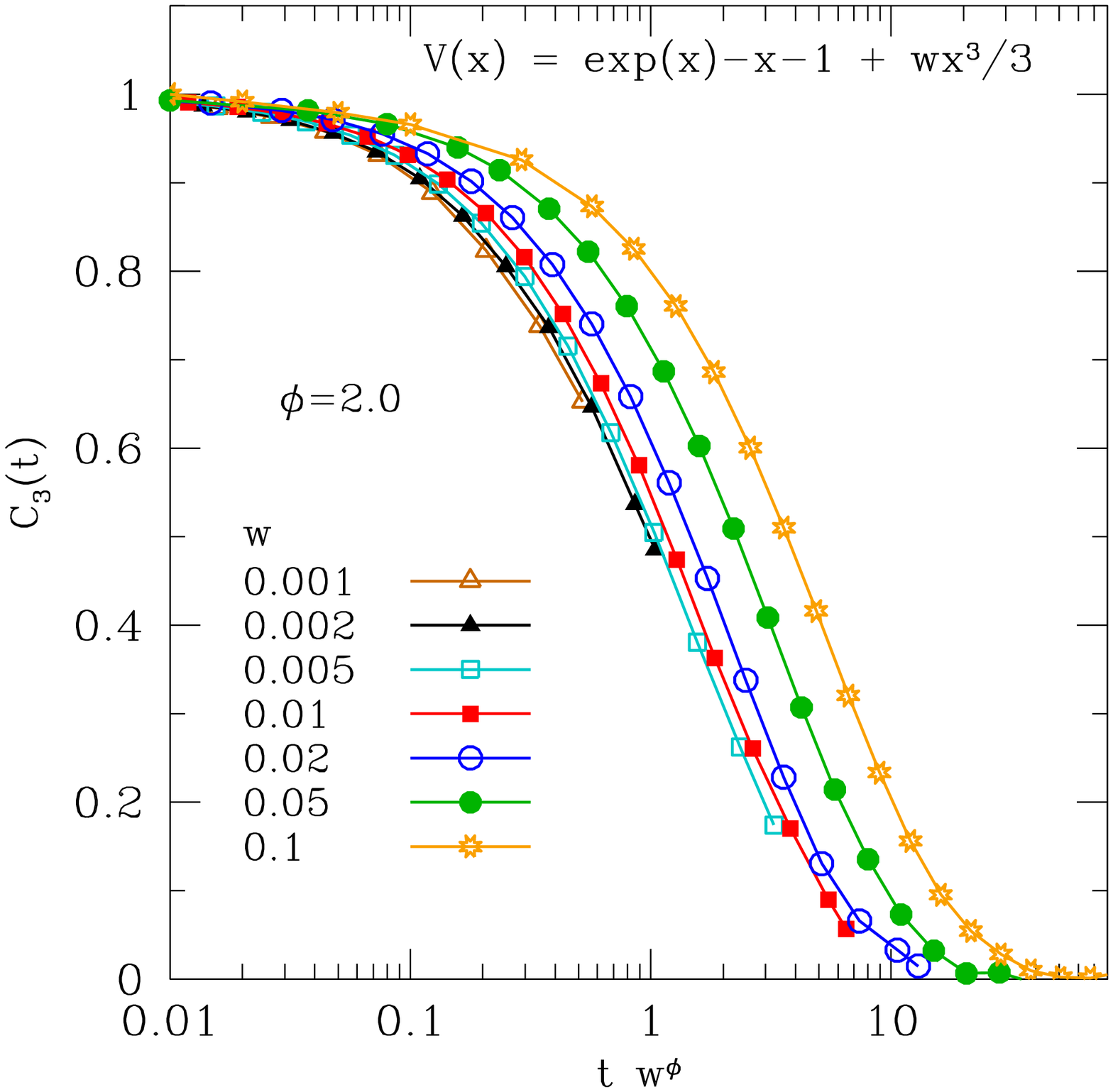}
\caption{Same as Fig.~\ref{J3_tau_scale_w_1.9} but for $\phi = 2$, which is
the expected result as discussed in the text. It is plausible that this 
value of $\phi$ will give good data collapse in the
scaling limit $w \to 0, t \to \infty$. This asymptotic regime is reached only
for \textit{very} small values of $w$ (and correspondingly very long times)
indicating that corrections to scaling are large.
}
\label{J3_tau_scale_w_2.0}
\end{figure}

We compute the normalized correlation function of the conserved current $J_3$,
see Eq.~\eqref{J3}, defined by
\begin{equation}
C_3(t) = {\langle J_3(t_0) \, J_3(t_0 + t) \rangle  \over
\llabel{C3}
\langle \left(J_3\right)^2 \rangle} \, .
\end{equation}

In the absence of any perturbation
which breaks integrability,
$w$ or $v$ in Eq.~\eqref{pert},
$C_3(t)$ is equal to unity. It is also equal to unity at $t=0$ for any
Hamiltonian.  Figure
\ref{J3_tau_several_w} shows data for $C_3(t)$ for the cubic
perturbation in Eq.~\eqref{pert}, for several values of the strength of
the perturbation $w$.  As expected the $J_3$ correlation function 
decays to zero on a timescale which
decreases with increasing $w$.

We assume that the data fits the scaling form
\begin{equation}
C_3(t) = \widetilde{C}_3(t w^\phi) \, ,
\llabel{C3_scale}
\end{equation}
where $\phi$ is a crossover exponent indicating that $w$ is a ``relevant''
perturbation for the integrable model.
Figure \ref{J3_tau_scale_w_1.9} shows scaled data for the cubic perturbation in
Eq.~\eqref{pert}
assuming $\phi = 1.9$ which gives the best data collapse for the largest
sizes.

However, as shown in Fig.~\ref{J3_tau_scale_w_2.0} it is also possible
that the data for even larger sizes will collapse for $\phi = 2$.
The result $\phi = 2$ (in a quantum version)  follows from  Fermi's golden rule.  It is commonly encountered in quantum
integrable systems~\cite{quantum-cases}, where the current decays due to the
addition of a term $V$ in the Hamiltonian that destroy integrability. The
Fermi golden rule states that the decay rate of a state $i$ is
given, to lowest order, by
\begin{equation}
\frac{1}{\tau} = \frac{2 \pi}{\hbar} \sum_j |\langle i | V | j \rangle|^2 \delta(\varepsilon_j- \varepsilon_i).
\end{equation}
This may be interpreted as the decay rate of a current or a
quasiparticle that is infinitely long lived in the integrable case. Hence
we expect that an integrable current analogous to $J_n$, and its
correlations would decay in this fashion, whereby we expect $\phi=2$. 

We note that the data in Fig.~\ref{J3_tau_scale_w_2.0}
only collapses for very small values of $w$ and very
large times, indicating that ``corrections to scaling'' are large. 

\begin{figure}[!tbh]
\includegraphics[width=\figurewidth]{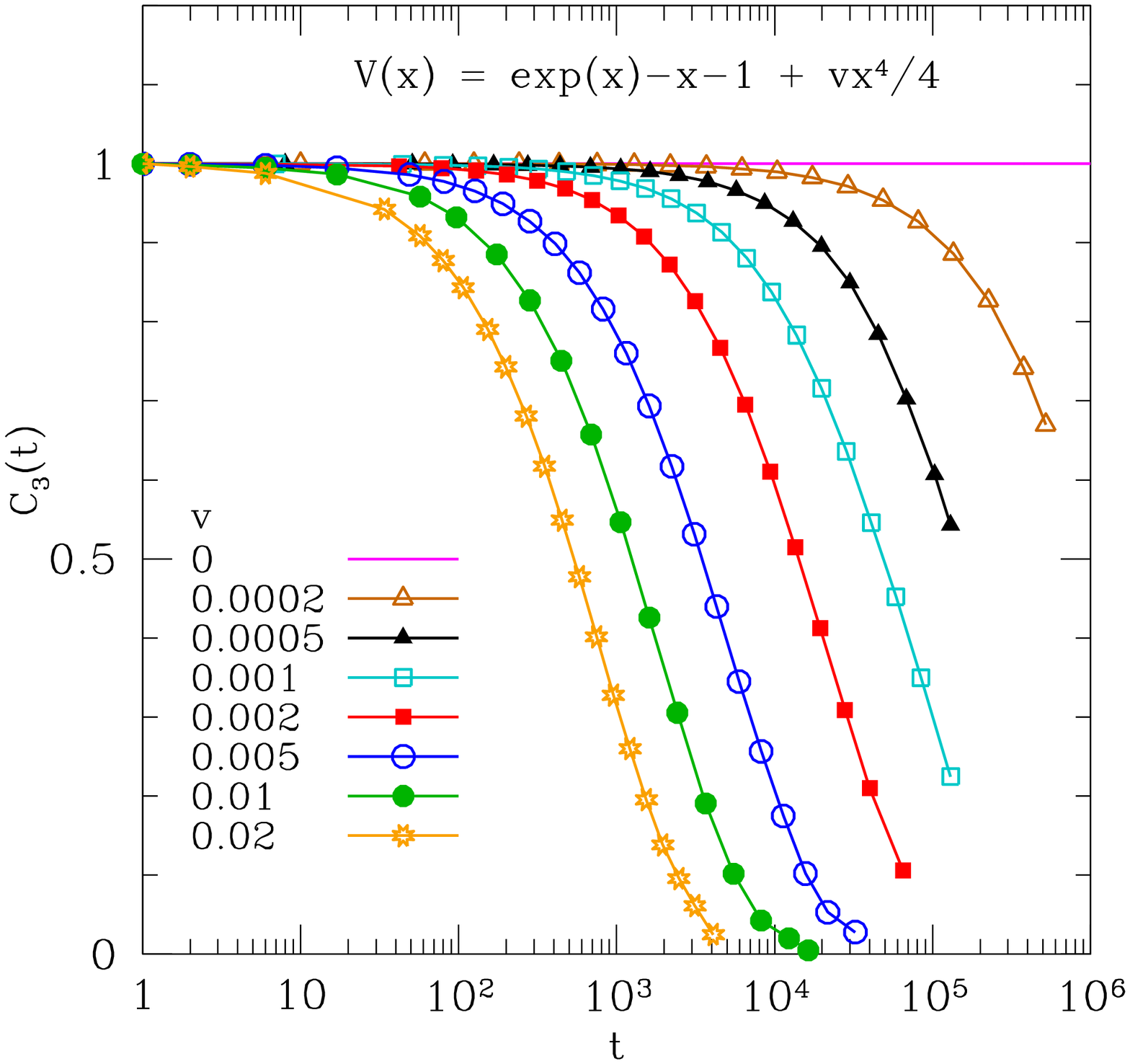}
\caption{The same as Fig.~\ref{J3_tau_several_w}
but for the quartic perturbation $v$ in Eq.~(\ref{pert}).
}
\llabel{J3_tau_several_u}
\end{figure}

\begin{figure}[!tbh]
\includegraphics[width=\figurewidth]{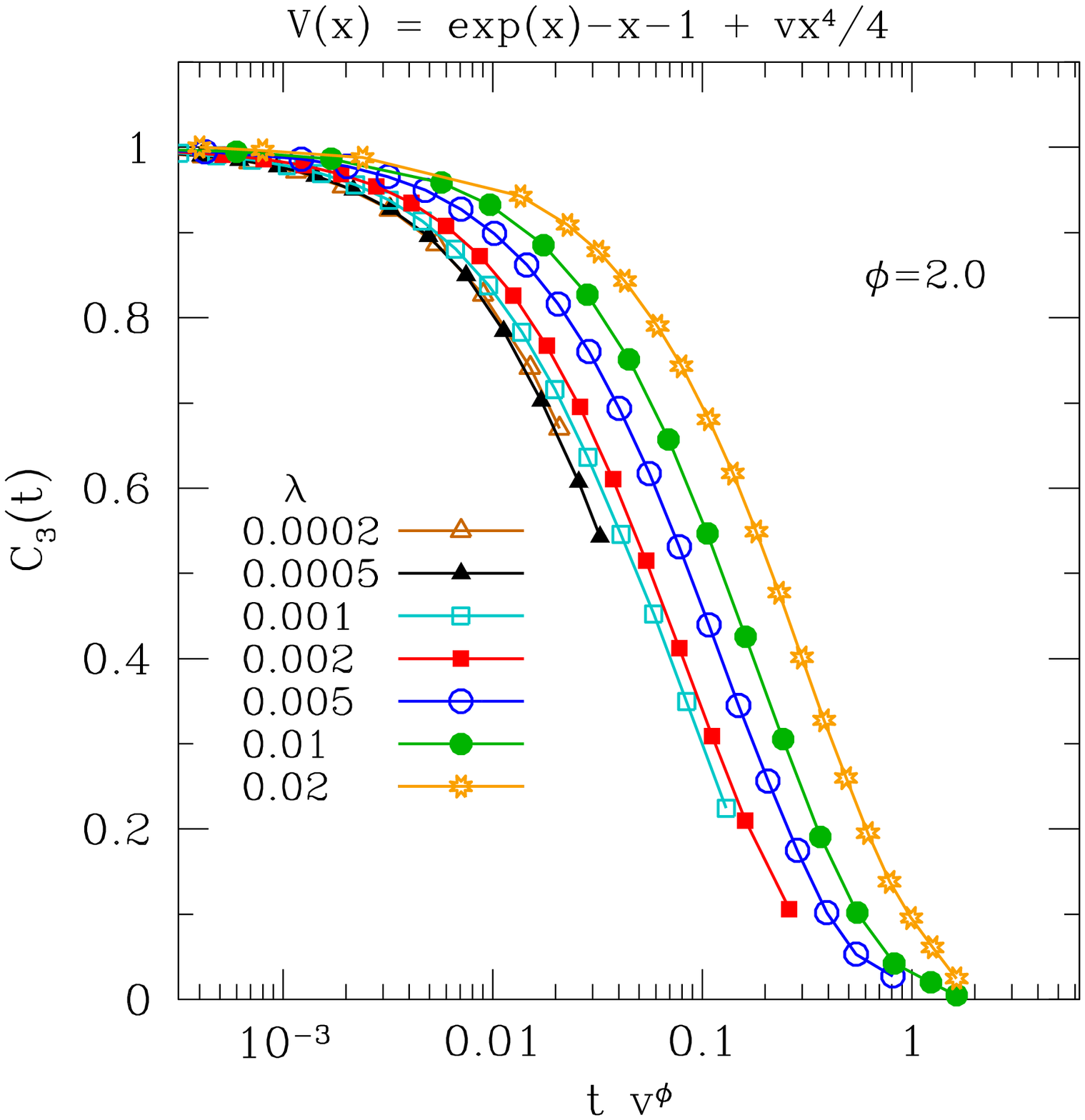}
\caption{A scaling plot with $\phi = 2$, as in
Fig.~\ref{J3_tau_scale_w_2.0},
but for the quartic perturbation $v$ in
Eq.~(\ref{pert}).
\label{J3_tau_scale_u_2.0}
}
\end{figure}

We have also studied the effect of the quartic perturbation in
Eq.~\eqref{pert} and show the unscaled data in
Fig.~\ref{J3_tau_several_u}.  We expect the same crossover exponent
$\phi$ for the quartic perturbation as for the cubic perturbation, and
indeed the data is consistent with the asymptotic exponent being $\phi =
2$ as shown in Fig.~\ref{J3_tau_scale_u_2.0}.

\section{Conclusions}

\llabel{conclusions}

In this paper we have shown that the result of Toda for the frequency of the
cnoidal waves of the Toda
lattice is only correct for weak anharmonicity or long wavelength. We
give a general expression for the frequency, as well as
results for the average kinetic energy and mean square
displacement in the Toda ring. The distinction between the dispersion relation of
Toda, Eq.~(\ref{disp-toda}),
and the one found here, Eq.~(\ref{disp-ring}),
is important only in the limit of large wave vectors and high
anharmonicity--in all other cases it is negligible.

In addition, we have discussed the conserved
currents of the 
Toda model in some detail. In particular, we showed numerically that the persistent
part of general currents can be expressed in terms of the conserved currents
according to Eq.~(\ref{expand-curr}),
provided one includes not only the Lax currents,
but also quadratic combinations of the Lax currents (which are, of
course, also conserved). Finally, we have studied the decay of the conserved
currents when a perturbation is added to the model which destroys
integrability. The timescale for decay is governed by a crossover exponent
$\phi$. Our numerical data
is consistent with the value $\phi = 2$, which can also
be obtained by Fermi golden rule-type arguments.

\acknowledgments
{We acknowledge support from the NSF through grants DMR-0706128 (BSS) and
DMR-0906366 (APY). We thank Professor Bill~Sutherland for 
sharing his unpublished notes on the periodic  case with us. 
His solution\cite{bill-unpub} has the same form that we report in
Eq.~(\ref{disp-ring}).}

\appendix
\section{Elliptic Theta Functions and Integrals 
\label{app-elliptic}}
The most convenient version of these functions
is given in Whittaker and Watson, Chapter XXI~\cite{ww}
and Abramowitz and Stegun~\cite{abramowitz-stegun}. We
note here the definitions of the required theta functions
and their Fourier series:
\begin{align}
q & =e^{- \pi K'/K}, & \theta_4(z) & = 1+2 \sum_{n=1}^\infty \ (-1)^n \
q^{n^2}\cos(2 n z) , \llabel{q} \\
K& \equiv K(m) = \int_0^{\frac{\pi}{2}} \ d x \ \frac{1}{\sqrt{1-m \sin^2 x}},
& \theta_1(z)& = 2 \ \sum_{n=0}^\infty \ (-1)^n q^{( n+\frac{1}{2})^2}\ \sin(2
n+1)z, \nn \\
K'&=K(1-m), &\theta_2(z)&= 2 \ \sum_{n=0}^\infty \  q^{( n+\frac{1}{2})^2}\
\cos(2 n+1)z, \nn \\
E&= \int_0^{\frac{\pi}{2}} \ d x \ {\sqrt{1-m \sin^2 x}}, &\theta_3(z)&=1+2
\sum_{n=1}^\infty \  \ q^{n^2}\cos(2 n z). 
\llabel{elliptic}
\end{align}
We note that for small anharmonicity,  all the elliptic functions can be expanded equivalently in terms of the parameter $m$ or the nome $q$. These expansions are related since we may expand
\beq
q=\frac{1}{16}m +\frac{1}{32} m^{2}+ \frac{21}{1024}m^{3}+\frac{31}{2048}m^{4}+O(m^{5}). \label{nome-expansion}
\eeq
 We note that as $m\to 1$, $K(m) \sim \log\{ \frac{4}{\sqrt{1-m}}\}$ and $E(m) \sim \frac{\pi}{2}$.
Further, as $m\to 1$, the theta functions are more easily calculated by using the Jacobi transformation\cite{ww}(p 475), where we display the parameter $m$
\beq
\theta_{n}(z|m)=c_{n} \ \frac{K}{K'} e^{-\{ z^{2} \frac{K}{\pi K'}\} } \ \theta_{n}( i \frac{ K}{K'} z|m), \llabel{jacobi-transform}
\eeq 
with $c_{1}=-i, \ \mbox{and the remaining  } \ c_{j}=1$.
For computing the soliton spectrum, we will need the expression for $n=1$
\beq
\theta_{1}(z|m)= 2 \ \frac{K}{K'} e^{-\{ z^{2} \frac{K}{\pi K'}\} } \ \sum_{n=0}^{\infty } (-1)^{n}e^{- \pi \frac{K}{K'}(n+\frac{1}{2})^{2}} \text{sinh} (2 n+1) \frac{K}{K'} z. \llabel{theta1-jacobi-transform}
\eeq 

\section{Toda's cnoidal wave frequency calculation, duality  and boundary
conditions
\label{app-toda}}

Consider N atoms in 1-d, with displacements $u_{1},u_{2 }\ldots u_{N}$, which we
call the bulk displacements. Additionally, there are two ``boundary'' atoms, atom 0
to the left of atom 1, and atom $N+1$ to the right of atom $N$. Here we do not
assume periodic boundary conditions but instead consider
two different types of boundary conditions:
\begin{enumerate}
\item
``clamped'' boundary conditions for which $u_0 = u_{N+1} = 0$, 
\item
and ``open'' boundary conditions for which, it turns out, we will need $u_0 = u_1$
and $u_{N+1} = u_N$.
\end{enumerate}
We will actually focus on ``mixed'' boundary conditions, i.e.~free at one end and
clamped at the other.

The interaction between two atoms is
represented by a nearest neighbor term $V(u_{j}-u_{j-1})$  with  a suitable
function $V$, the exponential interaction as in Eq.~(\ref{toda-H}), or a
harmonic term for comparison. 
The Lagrangian is
\beq
L = \frac{1}{2 M} \sum_{j=1}^{N} \dot{u}^{2}_{j} - \sum_{j=J_{l}}^{J_{r}} V(u_{j}-u_{j-1}), \label{app-1}
\eeq 
where $J_{l}=1\, (2)$ for clamped (open) boundary conditions
at the left, and $J_{r}=N+1\,(N)$
for a clamped (open) boundary conditions at the right.
Even for open boundary conditions, the
boundary sites do not have any kinetic energy since they are
``fictitious'', i.e. are simply there to impose the necessary boundary conditions.

We introduce the particle separations $r_{j}=u_{j}-u_{j-1}$ as new generalized
coordinates, these and their canonically conjugate momenta $s_{j}$ found below,  are the ``dual variables'' of Toda.
 It  follows that the inverse relations are
$u_{j}=\sum_{l=1}^{j}r_{l}$.  The kinetic energy is expressible in terms of
the time derivatives, $\dot{r}_{j}$. The canonically conjugate
``momenta'' to the $r_{j}$ are 
$s_{j}=\frac{\partial H}{\partial
\dot{r}_{j}}=M \sum_{l=j}^{N}\dot{u}_{l} $, whereby, for $1 \leq j \leq N-1$, we
get  $M \dot{u}_{j}= s_{j}-s_{j+1}$, and $\dot{s}_{N}= M \dot{u}_{N}$. Thus we
obtain the Hamiltonian in terms of the dual variables as
\beq
H= \frac{1}{2 M} \sum_{j=1}^{N-1}(s_{j}-s_{j+1})^{2 } + \frac{1}{2 M}
s^{2}_{N} + \sum_{j=J_{l}}^{J_{r}} V(r_{j}).
\eeq
Hamilton's equations of motion are therefore now
\beq
\dot{r}_{j}= \frac{1}{M} ( 2 s_{j}-s_{j-1}-s_{j+1}), \;\;\; \dot{s}_{j}= - \frac{\partial V(r_{j})}{\partial r_{j}}\;\;\;\text{for}\;\;2\leq j\leq N-1. \label{eom-toda-1}
\eeq

The case of a clamped right boundary is complicated to deal with so we will
always choose it to be open.
We therefore have $J_{R}=N$, and the
kinetic  energy term of the $N^{th}$ atom is written as $\frac{1}{2 M}
(s_{N}-s_{N+1})^{2}$ with $s_{N+1}=0$ as the boundary condition {\em at all
times}, so that the equation Eq.~(\ref{eom-toda-1}) is extended to $j=N$.

On the
left side boundary, the clamped case with $J_{l}=1$ can be dealt with easily
by extending Eq.~(\ref{eom-toda-1}) to $j=1$ by introducing an $s_{0}$ and
requiring that $s_{0}=s_{1}$ so Eq.~(\ref{eom-toda-1}) has the same form for
$j=1$ as for $j > 1$.
The open case on the left boundary  has a missing $V(r_{1})$ in
the potential energy so that $r_{1}$ is a cyclic coordinate (so $\ddot{r}_1=0$)
and we must
require $s_{1}=\mathrm{constant}$. 

For the case of the harmonic chain, these boundary
conditions are easily imposed on the  solution $e^{ \pm i \omega_{k}t} \cos(k
r_{j}+ \delta)$ and one reproduces the various integer and half integer
quantization of the wave vector $k$. We note that the solutions are not
traveling waves, but rather products of functions of space and time, since in
all cases here we must set the boundary variables to be time independent
constants.

For the Toda lattice, we choose $V(r_{j})= \frac{a}{b} \left( e^{-b r_{j} }-1+b r_{j}   \right)$ so that the equations of motion Eq.~(\ref{eom-toda-1}) are written for the typical case of left-clamped and right-open  boundaries
\beq
\frac{\ddot{s}_{j}}{a + \dot{s_{j}}}= \frac{b}{M} \left[ s_{{j-1}}+s_{j+1}- 2 s_{j}\right],\;\;1 \leq s_{j}\leq N, \;\;\;\text{and}\;\;\; s_{0}=s_{1}, \;\;\;s_{N+1}=0.\label{eom-toda-2}
\eeq         
The remarkable insight of Toda  in solving this equation, was apparently
inspired by seeing an addition identity of the Jacobean zeta function $Z[u]$
that
we met earlier in Eq.~(\ref{jacobi-zeta}). The salient features of this
function
are as follows:
periodicity $Z[u+2 K]= Z[u]$, parity $Z[-u]=-Z[u]$, and hence nodes at
$Z[0]=0=Z[K]$. The relevant identity is~\cite{ww-1}
\beq
Z[u+v]+Z[u-v]- 2 Z[u]= \frac{Z''[u]}{Z'[u]+ \frac{E}{K}-1 + \frac{1}{\text{sn}^{2}(v)}}. \label{zeta-addition}
\eeq 
Comparing Eq.~(\ref{eom-toda-2}) and Eq.~(\ref{zeta-addition}), one sees that these are very similar, {\em provided we make the hypothesis that  Eq.~(\ref{eom-toda-2}) should collapse to an ordinary differential-difference equation}. Thus we are obliged to combine the space and time dependence into  a single variable $\phi_{j}= k j - \omega t + \delta$, and this helps in solving the Eq.~(\ref{eom-toda-2}) for the bulk. We thus can only solve for traveling waves. However, it makes it impossible to satisfy the boundary conditions, since the latter involve time independent vanishing of certain constants. With Toda, {\em we will  ignore the boundary terms} and write down the solution for $s_{j}$ that is implied by Eq.~(\ref{zeta-addition}). With a scale factor $ \frac{\phi}{K} $ we relate $u$ and $\phi$ as   $u= \frac{\phi}{\pi} \ K$, so that increasing $u$ by  its natural periodicity $2 K$ winds the phase by $2 \pi$. Then we see that $\frac{d}{d t}= - \frac{K}{\pi} \omega \frac{d}{d u}$, and $j \to j+1$ increases $u \to u + \frac{k K}{\pi}$ whence $v=\frac{k K}{\pi}$. The mapping is complete with a scale factor relating  $s_{j}$ to $Z[u]$, and determining $\omega$ as a function of $k$  gives the two Toda  solutions: 
\barray
s_{j}(t)& =& \mp \frac{M}{b} \ \left(\frac{K \ \omega^{T}_{k}}{\pi}\right) \  Z[ \frac{K}{\pi} (k j \mp \omega^{T}_{k}t + \delta)] \label{toda-sol}  \\
\omega^{T}_{k}&=& \sqrt{\frac{a b }{M}} \ \frac{\pi}{K} \frac{1}{\sqrt{ \frac{E}{K}-1 + \frac{1}{\text{sn}^{2}(\frac{k K}{\pi})}} }. \label{disp-toda}
\earray

Using $M \dot{u}_{j}=s_{j}-s_{j+1}$  and the expression
Eq.~(\ref{jacobi-zeta}), and integrating once we find the displacement in
terms of the theta functions: \beq {u}_{j}= \frac{1}{b} \
\log\frac{\theta_{4}(\frac{1}{2} (k j \mp \omega^{T}_{k}t + \delta))
}{\theta_{4}(\frac{1}{2} (k (j+1) \mp \omega^{T}_{k}t + \delta))},
\label{toda-sol-2}
\eeq
in agreement with our Eq.~(\ref{sol-periodic}).

In Sec.~\ref{cnoidal} we showed that
the same functional form as
Eq.~(\ref{toda-sol-2}) describes oscillatory solutions with
periodic boundary conditions, but with a
different expression for the frequency, Eq.~(\ref{disp-ring}), in place of
Eq.~(\ref{toda-sol}). This situation requires a few clarifying remarks.

\begin{itemize}
\item
Eq.~(\ref{toda-sol}) is in a traveling wave form, and
since we cannot superpose two non linear waves, e.g. with the two signs of the
time dependence, this prevents us from satisfying the various boundary
conditions (clamped and open) discussed above. Thus the Toda dispersion
relation is
\textit{not the solution of the problem he starts with}. 
\item The spectra in
Eqs.~(\ref{disp-toda-2}) and (\ref{disp-ring}) are very close for small wave vectors,
 or for
small values of the parameter ``m'', i.e.~for weak anharmonicity.  Fig(\ref{fig-dispersion}) illustrates the two dispersions.
\item A reader might wonder if
Eqs.~(\ref{disp-toda}) and (\ref{disp-ring}) are not
actually identical, with the help of
some obscure identity, e.g.~in Ref.~\cite{ww}. However, this cannot be the
case for two reasons. One is mathematical.  While the expression in Eq.~(\ref{toda-sol}) is
doubly periodic in the complex $k$ plane, the expression in Eq.~(\ref{disp-ring}) is
not--the $\theta$ functions have periodicity factors attached for translations
along the imaginary axis. The other reason is physical. If the two expressions
were somehow identical, the potential energy average Eq.~(\ref{pot-average-1})
would be identically zero for the periodic system, which is impossible
since $V(x) \ge 0$ with the equality only at $x = 0$.
\item Toda is correct to 
take the functional form in Eq.~(\ref{toda-sol}) seriously. However for
periodic boundary conditions, the correct frequency of the cnoidal wave
is not the one in Toda's works\cite{toda}, but rather Eq.~(\ref{disp-ring}),
which
appears here for the first time, to the best of our knowledge. Toda's solution
is only correct in the limit of weak anharmonicity or long wavelength.
  
\end{itemize}

\section{Alternate derivation of the solution with periodic boundary
conditions.}
\label{alt-deriv}

In this appendix we give an alternative derivation of the dispersion relation
of the Toda ring in Eq.~(\ref{disp-ring}) by substituting
the Fourier series in
the second line of  Eq.~(\ref{toda-solution}) into Eq.~(\ref{eom-3}).
Taking the second derivative 
of this with respect to $\phi$ we obtain

\beq
 \text{LHS Eq.~(\ref{eom-3})} = 4 \ \overline{\omega}^{2}_{k} \
\sum_{n=1}^\infty n \ \frac{q^n}{1-q^{2n}}  \ \sin n (\phi- \frac{k}{2}
) \  \sin n \frac{k}{2}. 
\llabel{series-lhs}
\eeq
To obtain the expansion of the RHS of Eq.~(\ref{eom-3})
we start with Eq.~(\ref{toda-solution}) and use the Fourier expansion of $\text{sn}^2$
due to Jacobi~\cite{ww-2} (again!)
\begin{equation}
\text{sn}^2( \frac{ 2K}{\pi} x) =\frac{1}{m} \{ 1- \frac{E}{K} \}
- \frac{2 \pi^2}{m K^2} \sum_{n=1}^\infty n \frac{q^n}{1-q^{2n}}\ \cos 2 n x.
\llabel{jacobi-1}
\end{equation}
Thus we find
\begin{eqnarray}
\text{RHS Eq.~(\ref{eom-3})}&=& \frac{4
\pi^2}{\sqrt{m} K^2} \ \frac{\theta^2_1(\frac{k}{2})}{\theta^2_4(0)} \
\sum_{n=1}^\infty n \ \frac{q^n}{1-q^{2n}}  \ \sin n (\phi- \frac{k}{2}
) \  \sin n \frac{k}{2}. 
\llabel{series-rhs}
\end{eqnarray}
The series in Eqs.~(\ref{series-lhs}) and (\ref{series-rhs})
are seen to be identical, with the choice of the dispersion in
Eq.~(\ref{disp-ring}).

\section{ Potential energy of the Toda ring
\llabel{potential-energy} }

Let us rewrite the potential energy term Eq.~(\ref{pot-1}), with $\phi$
representing any one phase factor; 

\begin{eqnarray}
e^{b u_n-b u_{n+1}}& =& e^{d_k(\phi)- d_k(\phi+k)}= \frac{1}{\theta^2_4(0)}
\left\{ \theta^2_4(\frac{k}{2}) - \theta^2_1(\frac{k}{2}) \ 
\frac{\theta^2_1(\frac{\phi}{2})}{\theta^2_4(\frac{\phi}{2})} \right\},\nn \\
&=&\frac{1}{\sqrt{m}}  \frac{\theta_1^2(\frac{k}{2})}{\theta_4^2(0)} 
\left\{ \ \frac{1}{\text{sn}^2(\frac {k  K}{\pi})} - m  \
\text{sn}^2(\frac{K}{\pi}  \phi )  \right \}, \nn \\
&=& \bar{\omega}_k^2 \frac{K^2}{\pi^2} \ 
\left\{\frac{1}{\text{sn}^2(\frac {k  K}{\pi})} - m \
\text{sn}^2(\frac{K}{\pi}\phi)  \right \} . \llabel{pot-11}
\end{eqnarray}
We  used the two  dispersion relations for the chain and the ring,
Eq.~(\ref{disp-ring}), and Eq.~(\ref{jacobi-constants}) to proceed in this
equation. 

We now recall $\phi= \omega_k t$ and average this expression over a
single cycle in time, (i.e. $\int_0^{T_k} \cdots \frac{d t}{T_k}$ where
$T_k= \frac{2 \pi}{\omega_k}$), or $u= \frac{K}{\pi} \phi$, with $0 \leq
u  \leq 2K$. We ignore the site index, since each atom  has the same
average value. We use
$$  \frac{1}{2 K} \ \int_0^{2 K} \ du  \
\text{sn}^2(u) = \frac{1}{m} \left(\frac{E}{K} -1 \right) ,$$
so that the time average  of this term can be written using
the expression Eq.~(\ref{disp-toda}) as:
\beq
\overline{ e^{d_k(\phi)- d_k(\phi+k)}}=  \left(\frac{\overline{\omega}_{k}}{\overline{\omega}^{T}_{k}}\right)^{2},
\eeq
and thus the potential energy average over a cycle is
\beq
\overline{  \text{PE}}= N\, \frac{a}{b}\, \left[  \left(\frac{\overline{\omega}_{k}}{\overline{\omega}^{T}_{k}}\right)^{2}-1\right].\label{pot-average-1}
\eeq


\begin{thebibliography}{99}
\bibitem{fnote} Eq.~(\ref{jacobi-zeta-1})   is well known, e.g. see
Eq.~(17.4.38) of Ref.~\cite{abramowitz-stegun}.
The useful identity Eq.~(\ref{zeta}) is however not listed in literature as far as we could find, and it seems worth stating its derivation. 
 Here the Poisson summation formula is very useful, and gives Eq.~(\ref{zeta}) from Eq.~(\ref{jacobi-zeta-1}) in a few lines. We first write Eq.~(\ref{jacobi-zeta-1}) after adding and subtracting the $n=0$ term as:
$$Z(u)= - \frac{ \pi u}{ 2 K K'} + \frac{\pi}{2 K} \sum_{{n= -\infty}}^{{\infty}} \frac{\sin \frac{ n \pi u}{K}}{\sinh \frac{ n \pi K'}{K}}.$$
Since this is in the form of a Fourier ``sine-series'', we  use the ``sine-transforms'' version of the  Poisson summation formula:
$$\sum_{n=-\infty}^{\infty} g(n) \sin{  \frac{ n u}{u_{0}}}= u_{0}\sum_{\nu=-\infty}^{\infty}\hat{g}(u - 2 \pi u_{0}\nu), \;\;\;\mbox{where}\;\;\hat{g}(x)= \int_{-\infty}^{\infty} dy \ \sin(x y) \ g(u_{0} y),$$
with $u_{0}= K/\pi$ and $g(n)= \frac{\pi}{ 2 K} \frac{1}{\sinh( \pi K' n/K)}$, yielding the required result. Similarly Eq.~(\ref{toda-solution}) for the displacement is not listed in literature. It may be obtained by a similar analysis as the above series, or alternately  from  Eq.~(\ref{zeta})  by an integration with respect to the parameter $\phi'$ as given in the first line.
\end{thebibliography}
\end{document}